\documentclass[
 reprint,
nofootinbib,
 amsmath,amssymb,
 aps,
 prb,
]{revtex4-2}

\usepackage{graphicx}
\usepackage{dcolumn}
\usepackage{bm}
\usepackage{float}
\usepackage[usenames]{color}
\usepackage[colorlinks,bookmarks]{hyperref}
\usepackage[capitalise]{cleveref}
\usepackage[caption=false]{subfig}
\usepackage{multirow}
\definecolor{linkblue}{rgb}{0,0,0.8}
\definecolor{linkgreen}{rgb}{0,0.5,0}

\hypersetup{pdfpagemode=UseNone, pdfstartview=FitH, linkcolor=linkblue, %
            citecolor=linkgreen, urlcolor=linkblue}

\begin{document}

\title{A Multipolar Approach to Sliding Ferroelectricity} 

\author{Matthew Dykes${}^{1}$}
\affiliation{ ${}^1$Department of Physics, Cornell University, Ithaca, NY, 14853, USA}

\begin{abstract}

Traditional theoretical treatments of ferroelectricity do not straightforwardly extend to sliding ferroelectrics, which are increasingly-studied layered materials where a switchable electrical polarization is controlled by two-dimensional relative motion of the stacked layers. Therefore, in-depth analyses of the underlying processes which dictate their polarization behavior remain challenging. In this paper, we present a comprehensive approach for identifying the symmetry-adapted microscopic parameters which are responsible for driving the emergence of this polarization. First, we outline our approach, which appeals to group theory arguments and the distortion of Wannier orbital densities to connect macroscopic symmetries to the microscopic electronic distortions which dictate the appearance of ferroelectricity. Then, we illustrate this process by using density functional theory to apply our strategy to honeycomb bilayer systems, including hexagonal boron nitride. In this way, we find that combinations of dipole-like and quadrupole-like distortions of lone pair electron orbitals control electronic reorganization, and by extension, ferroelectricity in such systems.

\end{abstract}

\maketitle

\section{Introduction}
\label{sec:1}
Ferroelectrics are materials which possess switchable spontaneous electrical polarizations, making them ubiquitous components in electronic devices. As the demand for miniaturized, high component-density electronics has grown, increasing research focus has been placed upon 2D ferroelectric materials. In this area, sliding ferroelectrics have shown great promise. First proposed in 2017, sliding ferroelectricity arises in 2D layered van der Waals (vdW) systems where weak interlayer bonding allows neighboring layers to slide against each other along an in-plane (IP) direction, in the process generating a switchable out-of-plane (OOP) electrical polarization  $p_{\perp}$~\cite{li2017,wang2023}. 

In general, sliding ferroelectrics share numerous advantageous characteristics: resistance to depolarization fields, ultralow switching barriers, tunable band structures, thermal/mechanical stability and atomically sharp, clean interfaces~\cite{wu2021,zhang2023,s.li2024}. These properties translate straightforwardly to potential applications in nonvolatile memory storage~\cite{zhang2023,s.li2024,bian2023,yasuda2024}, neuromorphic computing~\cite{s.li2024,x.li2024,jin2022}, optoelectronics~\cite{s.li2024,dai2019,sun2022,xiao2022}, electromechanical systems~\cite{dai2019,xu2022,qi2021} and flexible electronics~\cite{zhang2023,jia2022}. In addition, some sliding ferroelectric systems can individually display intriguing physical phenomena, including ferroelectric metallicity~\cite{fei2018,yang2018}, superconductivity~\cite{jindal2023}, Moiré ferroelectricity~\cite{li2017,yasuda2021,viznerstern2021}, magnetoelectric coupling~\cite{k.liu2023,yu2024} and multiple polarization states~\cite{meng2022}. Due to this expanding research interest, the number of experimentally-verified sliding ferroelectrics has grown to include BN~\cite{yasuda2021,viznerstern2021}, MnPS$_3$~\cite{weng2025} $\gamma$-InSe~\cite{liang2025} and several transition metal dichalcogenides~\cite{meng2022,wang2022,wan2022,li2023,fei2018,sharma2019,ran2024}, with many more predicted theoretically. 

\begin{figure*}
	\centering
	\subfloat{%
		\begin{minipage}{\textwidth}
			\includegraphics[width=\textwidth]{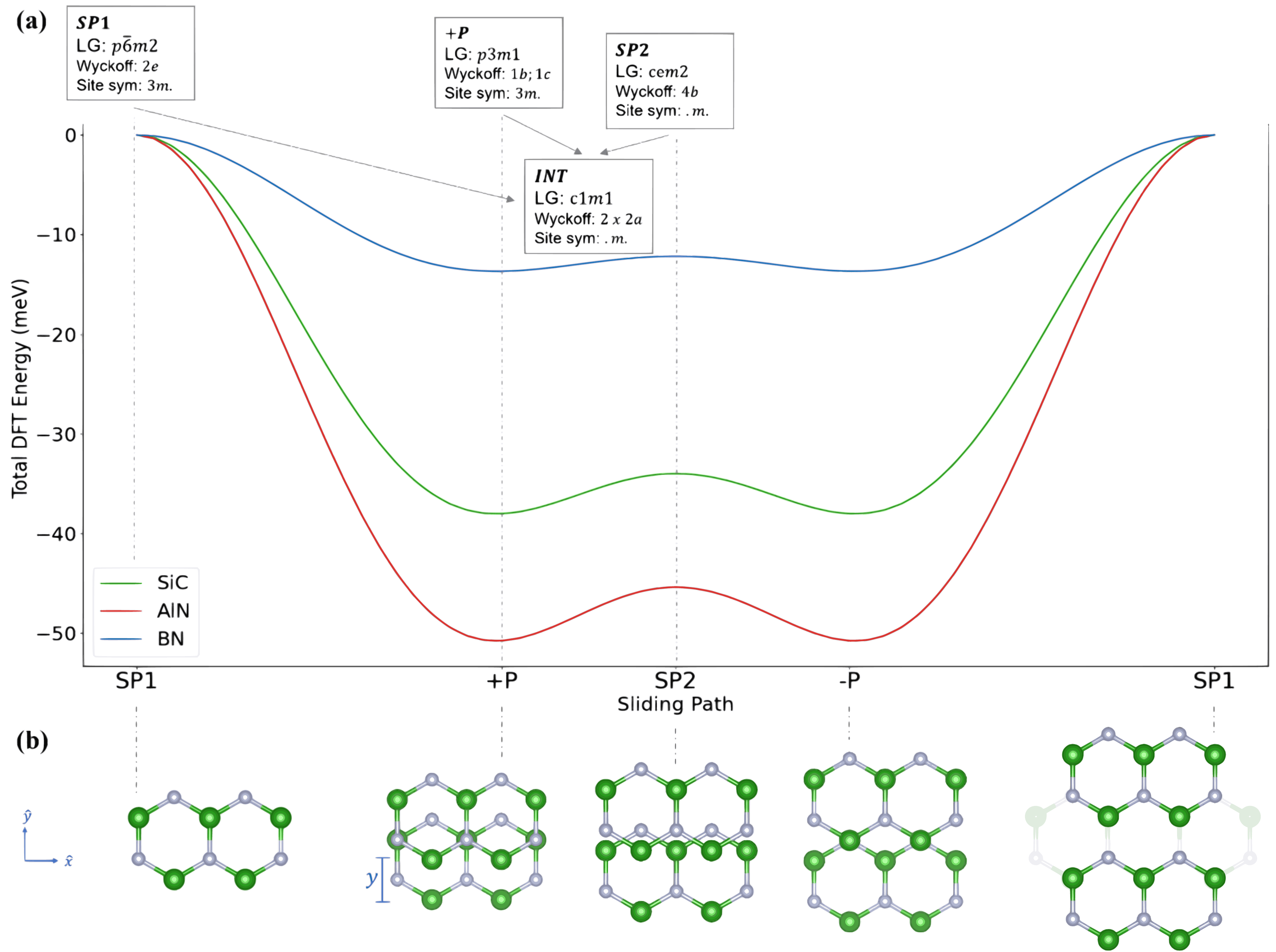}
		\end{minipage}%
	\label{Fig:1a}}
	\subfloat{%
		\begin{minipage}{0\textwidth}
			\includegraphics[width=\textwidth]{well_diagram-eps-converted-to.pdf}
		\end{minipage}%
	\label{Fig:1b}}
	\caption{Evolution of the vdW bilayer during sliding. a) Energy per unit cell, calculated using density functional theory (DFT), across one period of sliding, with colors corresponding to different atomic species of the $BA$ formula unit. Green spheres correspond to $A$ sites (Si, B, Al) and gray are $B$ sites (C, N).  The stacking arrangements of the higher-symmetry $SP1$, $\pm P$ and $SP2$ structures are shown in (b), where a small number of unit cells are shown in each layer. Boxes in (a) list the crystallographic groups of all distinct symmetry structures encountered along the period, as well as the Wyckoff positions and site symmetries of the $A$ sites (the $B$ sites always share the same site symmetries and often the same Wyckoff positions). Arrows indicate group-subgroup relations among these structures.}
\end{figure*} 

Despite the clear interest and excitement surrounding sliding ferroelectricity, we still lack a thorough exploration of the underlying mechanisms that ultimately cause it because traditional theoretical treatments used to understand ferroelectricity do not completely extend to sliding ferroelectrics. In typical structural ferroelectrics, the emergence of a polarization is driven by collective ionic displacements and bond distortions in the direction of the polarization, either directly (for proper ferroelectrics) or indirectly (for improper ferroelectrics) resulting from the parent phase losing stability with respect to a symmetry-adapted distortion mode. This collective distortion transforms like a partner function carrying the irreducible representation (irrep) that defines the symmetry reduction from the parent to the low-symmetry ferroelectric phase. Therefore, the microscopic atomic and bonding changes driving the transition are directly tied to the symmetry properties which define it at the level of group theory. 

Although some sliding ferroelectrics involve ionic displacements in addition to sliding~\cite{wang2023,xu2022,ma2021}, most consist only of a rigid sliding of vdW layers against each other, unaccompanied by OOP ionic displacements or covalent bond distortions. Thus, the polarization is completely electronic in origin, driven instead by distortions of the electron density~\cite{liu2019,rogee2022,zhong2021,yang2023b,wan2022,li2022,yang2023a,yang2018}. Despite this difference, sliding ferroelectrics are nonetheless driven by a smooth, symmetry-lowering transition. Therefore, whereas in most structural ferroelectrics we can explain polarization by identifying symmetry-adapted subsets of atomic displacements and bond distortions that are consistent with the symmetry lowering, in sliding ferroelectrics we must by analogy be able to explain polarization by identifying symmetry-adapted subsets of \textit{electron cloud distortions} which are consistent with the symmetry lowering. 

To our knowledge, sliding ferroelectric literature still lacks any discussion regarding a symmetry-adapted reordering of the electron density because this idea does not apply straightforwardly to sliding systems. Naively, a simple comparison between the electron densities of the initial nonpolar parent state and final polar state should highlight which changes in the density are consistent with the lowering of symmetry. Such a comparison between polar and nonpolar states is warranted if, as in traditional ferroelectrics, there exists a  symmetry-adapted distortion mode that directly connects both structures. Then, differences between the structures can be explained by scaling the magnitude of this mode. 

However, sliding ferroelectricity is an example of a reconstructive phase transition, in which no group-subgroup relation exists between the parent and polar states. This means that the initial nonpolar structure cannot distort directly into the final polar structure; instead they are linked by distortions to intermediate states of a lower, common-subgroup symmetry~\cite{lin2024,guo2024,bennett2023}.  Therefore, a simple comparison between the electron densities of the isolated initial and final states of the sliding transition cannot shed light on a symmetry-adapted density distortion which links them together.

Instead, it's necessary to study the full range of structures encountered during the evolution of the initial nonpolar state to the final polar state via intermediate structures. Such a treatment would capture the essential structural and symmetry changes that characterize sliding motion and therefore dictate what types of density distortions can occur. Focusing on bilayer systems, if we label any intermediate structure by $s$, the distance that both layers have slid with respect to one another compared to their initial configuration, the purpose of this paper is to devise an approach which explains the origin of polarization in sliding ferroelectrics by identifying symmetry-adapted components of local electronic distortions across a series of $s$, consistent with the global changes that define the transition. 

To inform this new approach, it's instructive to review current literature viewpoints regarding the origins of sliding ferroelectricity, which can be split into two main schools of thought. On one hand, some studies mainly concern the global symmetry changes that occur during, as well as accompanying energetic and polarization landscapes~\cite{lin2024,guo2024,bennett2023,yang2023b}, thereby providing an understanding of the general trends and transformation properties of materials tensors that arise and vary during sliding. On the other hand, different studies have addressed the physical electron cloud distortion aspect by computing differential charge densities (DCDs) $\delta n(\mathbf{r})$~\cite{liu2019,rogee2022,zhong2021,yang2023b,wan2022,jain2025}, equal to the difference in electron density between the polar stacking configuration and individual monolayers stacked in a nonpolar configuration. This approach is often accompanied by a localized decomposition of the electron density, either into atomic layer components via Bader or Hirshfeld analysis~\cite{yang2018,zhong2021,li2022,yang2023a,yang2023b}, or orbitals via maximally-localized Wannier functions (MLWFs)~\cite{wang2023,yasuda2021,k.liu2023,m.liu2023,ma2021}, to track how charge shifts between polar and nonpolar configurations. In all, the DCD/local decomposition picture demonstrates that the emergent polarization is due to an uneven OOP redistribution of electron density. 

While both schools of thought, the ``global" picture and the DCD picture, variously discuss the full evolution of macroscopic properties along the sliding pathway or microscopic distortions of the electron cloud, no approach combines both ideas. However, we require both points of view for our proposed approach which would combine symmetry evolution with microscopic electron cloud distortions. Both pictures are also needed for practical device design: the former provides trends which give insight into how a device would actually behave during switching, while the latter highlights the physical mechanism controlling polarization and allows us to inform device designs to leverage this mechanism.  

Therefore in this paper, we combine aspects of both viewpoints, while also expanding upon them, to present an approach for studying symmetry-adapted components of electron density distortions as a function of sliding distance between the nonpolar and polar bilayer structures. This combined approach will clarify the primary physical mechanism that drives the appearance of polarization during sliding ferroelectricity. 

We can separate our approach into three broad segments. First, we use global symmetry considerations to gain insight into the symmetry transformation properties that must be obeyed by this mechanism along the sliding pathway. Second, we use MLWFs as a localized basis to study DCDs as a function of sliding distance $s$. We propose strategies to both determine which MLWF orbitals are most relevant in contributing to polarization and study their evolution as the transition proceeds. Third, we devise a way to combine the distortion picture with the macroscopic symmetry picture in order to determine which combinations of relevant distortion components from the second step exhibit the transformation properties determined in the first step and thus represent the primary electronic response underlying the sliding ferroelectric phenomenon. Finally, we elucidate these points by using density functional theory (DFT) calculations to apply our approach to the example of honeycomb homobilayer systems with formula unit $BA$, where $A$ denotes the atom of greater electronegativity. 

\section{Our Approach}
\subsection{Global Symmetry Considerations of the Sliding Ferroelectric Landscape} 
\label{sec:f1}

In order to determine the underlying mechanism responsible for polarization, we begin by using group theory in conjunction with DFT calculations across a series of simulated structures to gain insight into the energetic, polarization and symmetry trends of the system versus sliding distance. We then use this insight to determine which symmetry changes are characteristic of sliding and which abstract transformation properties a hypothetical symmetry-adapted driving mechanism would need to obey. Then, as a consequence of symmetry, the physical electronic changes which dictate the appearance of polarization must also share these transformation properties. To be clear, by ``polarization mechanism", we mean a symmetry-adapted change in the electron density which exhibits the system's primary electronic response to sliding and thereby drives charge reorganization. This density change may itself be polar, or may couple to another object that transforms like a polar vector.
	
We idealize sliding motion as unidirectional, taking place along a high symmetry direction among rigid layers that don't rotate with respect to each other. In this approximation, sliding is periodic; a large enough relative displacement of the layers will eventually return the system to its original configuration. To fully understand materials properties trends and the conditions that periodicity imposes upon them, we must study the evolution of the system across one complete “period” of sliding motion, so we begin by laying out the energetic and symmetry trends that generally unfold over one period. 

Over the course of one period, the system passes continuously through several different stacking configurations and symmetry groups. Previous works have shown that in most sliding ferroelectrics, the energy landscape over one period resembles the double well encountered in traditional ferroelectrics~\cite{k.liu2023,lin2024,guo2024,bennett2023}. \cref{Fig:1a} shows this landscape for our honeycomb bilayer systems, computed in~\cref{sec:a1}, with all the different symmetries encountered along the period indicated. 

For our reference structure defining the $s=0$ state, we use the stacking configuration which places the atoms in the top layer directly above their counterparts in the bottom layer, as shown at the left of~\cref{Fig:1b}. This saddle point structure, which we refer to as $SP1$, belongs to subperiodic layer group $\mathcal{L}_{SP1}$ and is the highest-symmetry state and energetic maximum of the sliding period. As the system executes one period of sliding, it passes through three other distinct high-symmetry states: two energetic minima, corresponding to the polar states, and a smaller local energetic maximum separating them. We label these states $\pm P$ and $SP2$, with layer groups $\mathcal{L}_P$ and $\mathcal{L}_{SP2}$, respectively. All other intermediate structures encountered along the period belong to layer group $\mathcal{L}_{INT}$, which is a subgroup common to the three higher-symmetry groups. 

\begin{figure}
\centering
    \subfloat{%
        \begin{minipage}{0.48\textwidth}
            \includegraphics[width=\textwidth]{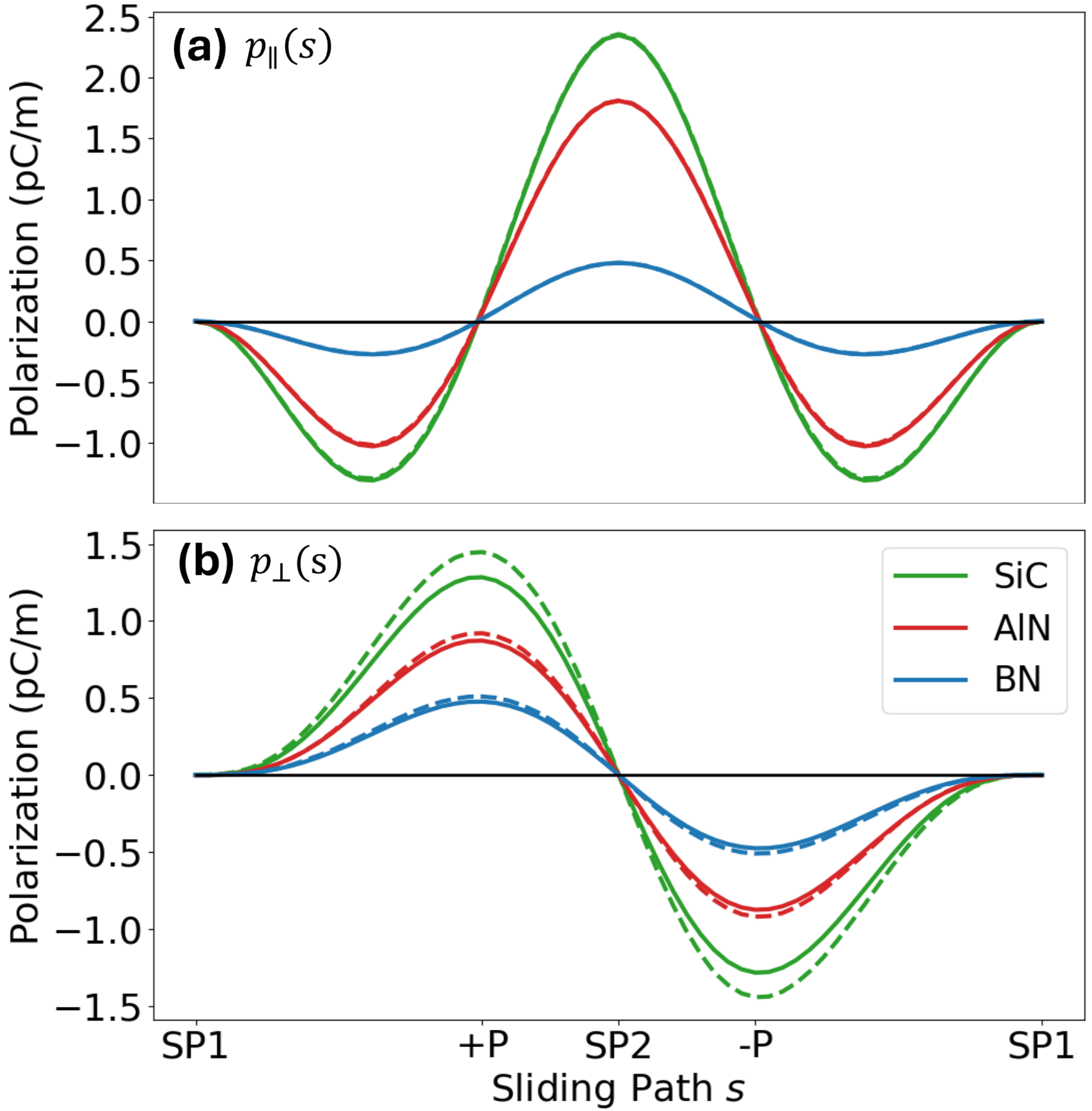}
        \end{minipage}%
    \label{Fig:2a}}
    \subfloat{%
        \begin{minipage}{0\textwidth}
            \includegraphics[width=\textwidth]{fig2-eps-converted-to.pdf}
        \end{minipage}%
    \label{Fig:2b}}
\caption{Polarization profiles of a) $p_{\parallel}$ and b) $p_{\perp}$ for  SiC, BN and AlN honeycomb bilayer systems, calculated across one sliding period. Solid lines display the results of Berry phase calculations and dashed are the results of maximally-localized Wannier functions (MLWF) calculations using~\cref{eqn:wannier_pol}.}
\end{figure}

It has been shown~\cite{bennett2023} that the functional form of the IP and OOP polarization components along the sliding period, $p_{\parallel}(s)$ and $p_{\perp}(s)$, can be derived using knowledge of the above symmetry groups alone (see~\cref{eq:7,eq:8} and Section SIV A~\cite{supp}). Examples of these profiles are depicted in~\cref{Fig:2a,Fig:2b}, computed for our honeycomb bilayer systems in~\cref{sec:a1}.

Similar to the treatment of traditional ferroelectrics, we postulate that polarization changes across the entire sliding period can be attributed to a single driving mechanism. In other words, electronic reordering throughout the period can be controlled by turning one master knob. Not only is this a sensible assumption since we've idealized sliding motion as unidirectional, but an underlying global ``knob" or continuous symmetry relationship between different states must exist due to compatibility constraints inherent in a periodic system.
 
As discussed in~\cref{sec:1}, sliding ferroelectricity is a reconstructive transition where the low-symmetry and  high-symmetry structures are bridged by an intermediate structure which they share as a common subgroup~\cite{capillas2007}. In fact,~\cref{Fig:1a} shows that a full period of sliding ferroelectricity consists of four reconstructive transitions: $SP1 \rightarrow INT \rightarrow +P$, $SP2 \rightarrow INT \rightarrow +P$, $SP2 \rightarrow INT \rightarrow -P$ and $SP1 \rightarrow INT \rightarrow -P$, where each transition is linked to the next by a shared endpoint. Therefore, similar to a symmetry study of phonon or electronic bandstructures, there must exist compatibility relations among all of the irreps and corresponding partner functions which define transitions among the four different groups encountered during the period~\cite{dresselhaus}. Thus, to grasp the essential symmetry changes which characterize the sliding transition, we need to analyze the four reconstructive transitions under the compatibility constraint. However, only the first two listed are independent by symmetry, so we focus solely on these.

Typically, the driving mechanism responsible for a ferroelectric transition from the high to low-symmetry phase is characterized as an abstract order parameter (OP) within the framework of Landau theory. Although Landau Theory doesn't apply to reconstructive transitions in which displacements are large, its ideas can still be applied to the immediate vicinity of saddle points where displacements are still small. So, for two independent reconstructive transitions, we can imagine prescribing two ``OPs," $OP_{SP1\rightarrow INT}$ and $OP_{SP2\rightarrow INT}$, each representing the spontaneous symmetry reduction from an unstable nonpolar structure to nearby intermediate structures via slight displacements. We could also identify an $OP_{+P\rightarrow INT}$ that induces the hypothetical $+P\rightarrow INT$ transition. This must exist by symmetry although from~\cref{Fig:1a} it is energetically forbidden. 

Once we identify the irreps and partners which carry these OPs, we can ``connect the dots" to interrelate them using compatibility constraints. Putting everything together, if our first goal is to prescribe an abstract macroscopic parameter $\boldsymbol{\eta}$ to our overarching polarization-driving distortion which specifies its magnitude and symmetry transformation properties, then compatibility enforces three conditions which it must satisfy: at $SP1$, $SP2$ and $+P$, $\boldsymbol{\eta}$ must behave like $OP_{SP1\rightarrow INT}$, $OP_{SP2\rightarrow INT}$ and $OP_{+P\rightarrow INT}$, respectively. In other words, $\boldsymbol{\eta}$ must transform as a partner of each of the irreps carrying the $SP1\rightarrow INT$, $SP2\rightarrow INT$ and $+P\rightarrow INT$ transitions. Thus, ultimately, the symmetry-adapted polarization-controlling parameter $\boldsymbol{\eta}$ must belong to the intersection of the three sets of irrep partners carrying each of $SP1\rightarrow INT$, $SP2\rightarrow INT$ and $+P\rightarrow INT$. 

As was the case for $p_{\perp}$ and $p_{\parallel}$, and as detailed in Section SIV A~\cite{supp}, once the transformation properties of $\boldsymbol{\eta}$ are known, we can use symmetry considerations across the sliding landscape to find a functional form $\boldsymbol{\eta}(s)$ which describes how $\boldsymbol{\eta}$ varies with sliding. Finally, the influence of the master electronic mechanism $\boldsymbol{\eta}$ on the appearance of $p_{\perp}$ and $p_{\parallel}$ can be revealed by studying the invariant free energy polynomials $\mathcal{F}(\boldsymbol{\eta},p_{\parallel},p_{\perp})$ that can be formed about the energetic extrema and noting the couplings between the polarizations and $\boldsymbol{\eta}$.

\subsection{MLWFs and Their Evolution During Sliding}
\label{sec:f2}

With a clear picture of the overall symmetry and evolution behavior of the abstract transition-defining parameter $\boldsymbol{\eta}$ in mind, our next step is to ascribe this behavior to something physical and concrete through an analysis of electron density changes. As stated in~\cref{sec:1}, we will need to compute the DCD as a function of $s$, $\delta n(\mathbf{r},s)$, to have any hope of relating microscopic distortions to trends across the sliding period. 

However, rather than treating the $\delta n(\mathbf{r},s)$ as monolithic objects, it's necessary to break them into local components because computing a simple total difference in the electron density is a coarse-graining procedure that cannot specify which, if any, components of the charge density are essential to the phase transition. For example, some components of $\delta n(\mathbf{r},s)$ may be indicative of a primary response to sliding, while others may be coupled secondary effects;
this distinction is essential in our goal of specifying a polarization-driving mechanism which captures the primary electronic response to sliding. 

We choose to decompose the electron density into contributions from MLWFs. In addition to those studies of sliding ferroelectricity that highlight the importance of MLWFs~\cite{wang2023,yasuda2021,k.liu2023,m.liu2023,ma2021}, some studies in bulk ferroelectrics also underscore the importance of instabilities in specific orbitals~\cite{smith2015,shen2019,jeong2021}. Although individual Wannier orbitals are gauge dependent, we fix the gauge using the MLWF prescription because MLWFs are heavily utilized and trusted within computational chemistry due to their ability to capture chemically-intuitive behavior which generally adheres to orbital and bonding theory~\cite{marzari2003,marzari2012}. Thus, they are an ideal basis for our goal of decomposing DCDs as a function of $s$ into chemically and energetically-intuitive components capable of discriminating between direct and indirect contributions toward the electronic response to sliding. 

As detailed in Section SIV B~\cite{supp}, the decomposition of a DCD into MLWF contributions is given by:
	\begin{gather}
	\label{eqn:dmlwfd}
	\delta n(\mathbf{r},s)=\sum_{i,\mathbf{R}}\delta n_{i,\mathbf{R}}(\mathbf{r},s) \\
	\delta n_{i,\mathbf{R}}(\mathbf{r},s) = |w^s_{i,\mathbf{R}}(\mathbf{r})|^2-|w^0_{i,\mathbf{R}}(\mathbf{r})|^2 \nonumber
\end{gather}
where $w^s_{i,\mathbf{R}}(\mathbf{r})$ is the $i^{th}$ MLWF in the unit cell associated with lattice vector $\mathbf{R}$, calculated for a bilayer with sliding distance $s$. In this way, the differential MLWF densities (DMDs) $\delta n_{i,\mathbf{R}}(\mathbf{r},s)$ that we've constructed each describe how much the component of the total electron density contributed by MLWF $w^s_{i,\mathbf{R}}(\mathbf{r})$ has changed after the bilayer has slid a distance $s$, compared to its value in the reference $SP1$ configuration. 

With DMDs, we can study how MLWF densities each microscopically distort as a sliding period is traversed, and a detailed analysis of their behavior will allow us to discern the primary subset of DMDs that are most relevant in capturing the direct electronic response to sliding and which will therefore be in physical correspondence with $\boldsymbol{\eta}$. To begin paring down the orbitals into a relevant set, we first discard those not belonging to the $\mathbf{R=0}$ home unit cell that we use for all our analysis, and then continue, if necessary, to remove others by appealing to previous results and chemical intuition, to be discussed in greater detail in~\cref{sec:a2}. 

Most importantly, we can also relate DMD distortion behavior with energetic intuition to further discriminate relevant orbitals. At the extrema of the~\cref{Fig:1a} energy landscape, the system is in equilibrium; the net force acting on MLWF densities is zero and is balanced among both interlayer and intralayer interactions with the other ions and MLWF densities. When the system slides away from the extrema, the forces become unbalanced, causing MLWF distortions. We can consider some MLWFs as being more important or essential to the electronic response to sliding if their strongest interactions in this unbalanced regime are interlayer as opposed to intralayer because, in this case, the overriding character of their distortions must be driven by sliding. Thus, such orbitals dictate how charge begins to reorganize as a direct result of sliding. If a MLWF density's intralayer interactions are strongest, then it plays a secondary role; it will respond to sliding, but indirectly as a result of interactions with the MLWF densities which respond directly.

To identify MLWF densities whose dominant interaction is interlayer, we study the energetic landscape of the problem. Since the layers are weakly interacting, we can reasonably approximate the interlayer interaction felt by an MLWF density in one layer, $L_1$, due to the other layer, $L_2$, as the application of the potential energy field of an isolated monolayer placed at $L_2$'s location. Then, when a slide of size $s$ occurs, the change in potential felt by the $L_1$ MLWF density is a lateral shift of this monolayer potential landscape by $s$, plus a change in the intralayer potential. If interlayer interactions dominate over intralayer ones, then when the system slides away from an extremum, the lateral shift of the $L_2$ potential landscape will lead the $L_1$ MLWF density to experience an unbalanced force and corresponding distortion roughly in the direction of steepest descent toward the minimum of the $L_2$ potential. Therefore, to rule out any other MLWFs as irrelevant, we must identify the minimum of the monolayer potential landscape and determine which DMDs show directionality consistent with its lateral shifting during sliding, as will be discussed in greater detail in~\cref{sec:a2}.

Through all the above considerations, we will arrive at some set $W$ of $N_W$ important DMDs that we will use for further analysis:
\begin{equation}
	\label{eqn:DMDset}
	W=\{\delta n_{I_1 \mathbf{0}},\delta n_{I_2 \mathbf{0}},...,\delta n_{I_{N_W} \mathbf{0}}\}\equiv \{\delta n_{I_1},\delta n_{I_2},...,\delta n_{I_{N_W}}\}
\end{equation}
where the reference to the $\mathbf{R=0}$ unit cell is dropped in all further analysis and $I_i$ is the index of the $i^{th}$ MLWF relevant for determining the primary response to sliding.

\subsection{Connecting the Local Distortion Picture to the Macroscopic Symmetry Picture}
\label{sec:f3}

\begin{figure*}
\includegraphics[scale=.195]{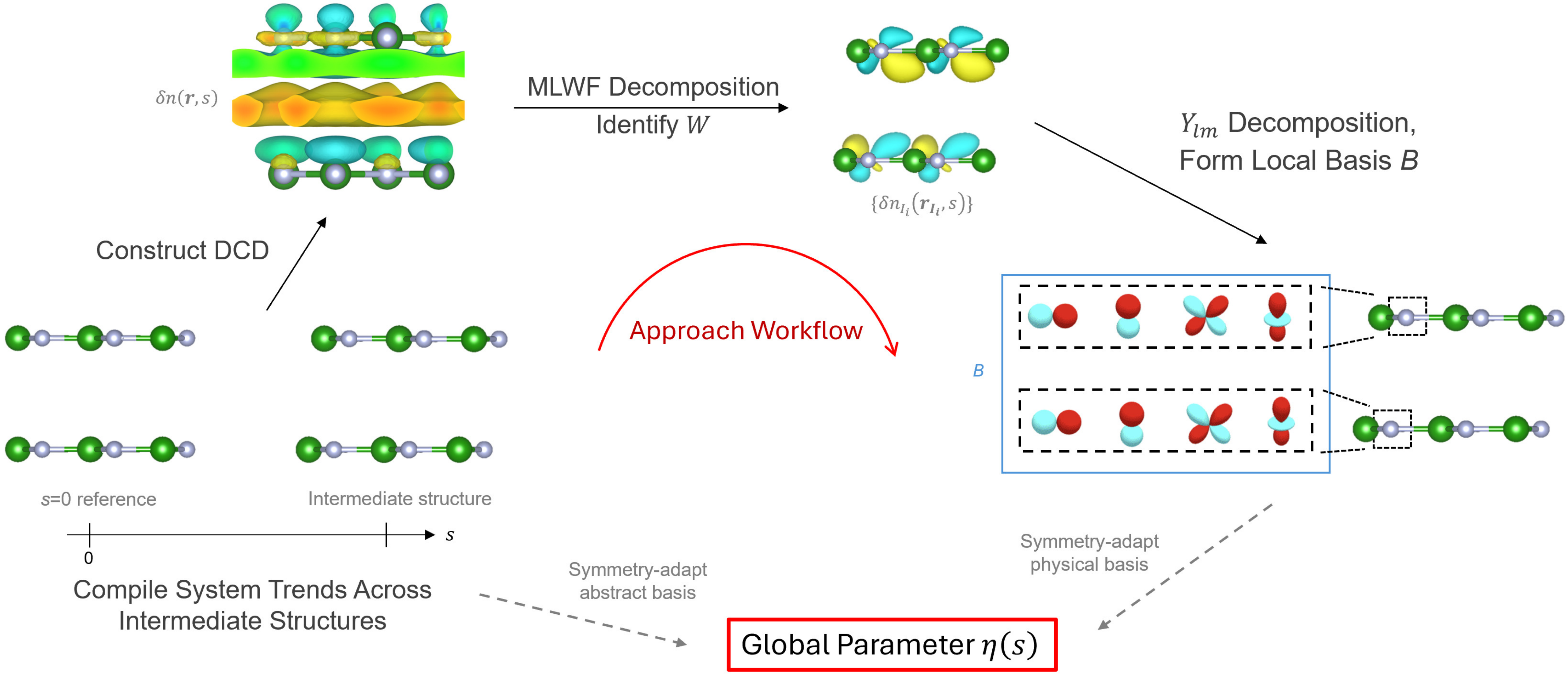}
\caption{An outline of our approach for evaluating the polarization-driving electron density distortion $\eta(s)$ for an intermediate structure with sliding distance $s$. After using macroscopic symmetry trends to gain intuition for the functional form of $\eta(s)$, we compute the differential charge density (DCD) $\delta n(\mathbf{r},s)$ and use an MLWF decomposition to eventually reduce it to a set $W$ of relevant orbital distortions $\{\delta n_{I_i}(\mathbf{r_{I_i}},s)\}$. Then, we decompose these distortions into symmetry-adaptable local components using $Y_{lm}$ functions and use them to formulate a new basis $B$ of relevant MLWF density distortion components. Finally, by applying global symmetry considerations to the functions in $B$, we construct a density distortion which shares the same transformation properties as $\eta(s)$, thereby linking the abstract symmetry properties which define the phase transition to an underlying physical object.}
\label{fig:outline}
\end{figure*}

After following both~\cref{sec:f1} to determine the overall symmetry properties of $\boldsymbol{\eta}$ and~\cref{sec:f2} to determine the set $W$ of DMDs which primarily contribute to it in real space, we can now devise a way to marry the former abstract picture with the latter physical picture. In short, we require a way of relating the DMDs in $W$ to the system's symmetries in order to achieve our goal of identifying a physical electron density object transforming like the abstract $\boldsymbol{\eta}$. Specifically, we seek to determine which components and linear combinations of DMDs form an electronic basis obeying the overall symmetry properties of $\boldsymbol{\eta}$ by first decomposing the DMDs in $W$ into symmetry-adapted components localized around each layer. Then, we find linear combinations of these components which share the transformation properties of $\boldsymbol{\eta}$.

Since DMDs are real, square integrable functions, the real spherical harmonics $Y_{lm}$ are suitable as a basis for this local decomposition~\cite{jackson}. Also, they are an ideal choice for a symmetry-adapted decomposition because they have definite symmetry transformation properties, which allow them to be used as partner functions to carry any irrep of any crystallographic point group (PG)~\cite{altmann}. 

The real spherical harmonic decomposition of a DMD in $W$ is given by 
\begin{equation}
	\label{eqn:ylm_decomp}
	\delta n_{I_i}(\mathbf{r_{I_i}},s)=\sum_{lm}{c_{lm}^{(I_i)}(r_{I_i},s)\times Y_{lm}(\theta_{I_i},\phi_{I_i})}.
\end{equation}
Since this decomposition is defined on the radius of a sphere, we set a spherical coordinate system $\mathbf{r_{I_i}}$ for each DMD $\delta n_{I_i}$, centered at the point in space about which $\delta n_{I_i}$ is generally largest in magnitude. As is shown in Fig. S1~\cite{supp}, this point can be clearly determined visually for our honeycomb bilayer systems. 

Since the sum in~\cref{eqn:ylm_decomp} is infinite, we need a practical way of restricting our focus only to $Y_{lm}$ components which contribute appreciably to the decomposition, as most will only contribute negligibly.  A sensible measure of the ``strength" $\mathcal{C}^{(I_i)}_{lm}(s)$ of each $Y_{lm}$ component in the decomposition would be how much each contributes to the $L^2$ norm of $\delta n_{I_i}(\mathbf{r_{I_i}},s)$, which is proportional to
\begin{equation}
	\label{eqn:spectral_strength}
	\mathcal{C}^{(I_i)}_{lm}(s)=\int |c_{lm}^{(i)}(r_{I_i},s)|^2 r_{I_i}^2 dr_{I_i} .
\end{equation}
By plotting $\mathcal{C}^{(I_i)}_{lm}(s)$ across $s$ for several $(l,m)$, we can determine which $Y_{lm}$ components contribute appreciably to the relevant DMDs. 

From this analysis, we will find a set $Y^{(I_i)}=\{Y_{l_1m_1}^{(I_i)},~Y_{l_2m_2}^{(I_i)},~...,~Y_{l_{N_{I_i}}m_{N_{I_i}}}^{(I_i)}\}$ of $N_{I_i}$ spherical harmonics which, in general, comprise the vast majority of the decomposition~\cref{eqn:ylm_decomp} for $\delta n_{I_i}$ across a range of $s$. Here, the $I_i$ superscript denotes the $Y_{lm}$ with its origin located at the origin of the $\mathbf{r}_{I_i}$ coordinate system. Thus, after reducing the problem to a set of $N_W$ primarily-relevant DMDs and $N_{I_i}$ primary spherical harmonic components for each DMD, we've effectively created a $\sum_{I_i=1}^{N_W} N_{I_i}$-sized basis $B$ reflecting local symmetry-adapted changes in the electron density which are most important in describing how charge reorganizes during sliding.

Finally, we can combine these locally-referenced basis functions through linear combination into functions which share the transformation properties ascribed to $\boldsymbol{\eta}$ via~\cref{sec:f1} across the entire bilayer system. For each distinct PG encountered as $s$ varies, the set $B$ constructed in the preceding step generally forms a basis which carries a reducible representation (rep) for the PG\footnote{As will be made clear in~\cref{sec:a3}, some discarded $Y_{lm}$'s may need to be added back to $B$ if the PG irrep is multidimensional.}. Then, we break these reps into irreps and pick out those for which $\boldsymbol{\eta}$ was determined to be a partner through the analysis of~\cref{sec:f1}. Finally, we use projection operators~\cite{dresselhaus} to determine which linear combinations of basis functions in the rep space $B$ carry the desired irrep. 

In summary, we have determined a basis set $B$ which describes relevant local distortions of the system's MLWF densities and used these basis functions to construct partner functions for the irreps which define the sliding transformation from a macroscopic perspective. Accordingly, these resulting global symmetry-adapted linear combinations of local spherical harmonics are valid functional expressions of $\boldsymbol{\eta}$ and represent a sliding ferroelectric's primary electronic response to sliding.

Since our final basis functions are stand-ins for $\boldsymbol{\eta}$, they can also be used to calculate $\boldsymbol{\eta}(s)$ if we can determine how the distortion components contributed by these basis functions vary with $s$. This will also provide an empirical way of confirming the functional form of $\boldsymbol{\eta}(s)$ determined previously by symmetry considerations alone. We need a scaling factor $\eta^{(I_i)}_{lm}$ that describes the ``amount of" $Y^{(I_i)}_{lm}$ in $\delta n_{I_i}$, which we can obtain by projecting out this component via the integral $\eta^{(I_i)}_{lm}(s)=\int\delta n_{I_i}(\mathbf{r_{I_i}},s)Y_{lm}(\theta_{I_i},\phi_{I_i})d^3\mathbf{r_i}$, which using~\cref{eqn:ylm_decomp}, gives
\begin{equation}
	\label{eqn:eta}
	\eta^{(I_i)}_{lm}(s)=\int c^{(I_i)}_{lm}(r_{I_i},s)r_{I_i}^2 dr_{I_i}.
\end{equation}  
The coefficients of partner functions in a linear combination also carry the irrep that the partner functions carry~\cite{toledano,landau}. Therefore, combining these $\eta^{(I_i)}_{lm}(s)$ coefficients in the same linear combinations as the corresponding $Y^{(I_i)}_{lm}$'s in the global basis functions determined in the previous step should give the smooth function $\boldsymbol{\eta}(s)$.

To summarize, our approach, which is outlined by the flow chart in~\cref{fig:outline}, elucidates the polarization mechanism in sliding ferroelectrics by first determining an abstract OP-like quantity $\boldsymbol{\eta}$ which captures the symmetry behavior of the sliding instability. Then we break down the system's electronic distortions during sliding into relevant local components and use $Y_{lm}$ decomposition and group theory arguments to transform them into an equivalent, physically meaningful realization of $\boldsymbol{\eta}$. In this way, we've plucked out the most important subset of electronic distortions which characterizes the electronic response to sliding and hence ultimately drives polarization, either directly or through coupling to another polar distortion. 

\section{Application the Honeycomb Bilayer Systems}

Our approach can be clarified by an application homobilayer vdW systems, in particular honeycomb bilayer SiC, BN and AlN structures. While the layered honeycomb structure is well-known in BN and is confirmed to host sliding ferroelectricity, theoretical and experimental results suggest that SiC and AlN can also stabilize in this structure in the ultrathin limit~\cite{freeman2006,lin2012,polley2023,tsipas2013}. The artificial stacking of two such layers these compounds is also expected to yield sliding ferroelectricity~\cite{wang2023}. These systems are ideal toy models for applying our approach due to their simplicity; they are all predicted to be nonmagnetic, wide band-gap semiconductors with few valence orbitals~\cite{sahin2009,ha2023,chettri2021,bacaksiz2015}. 

In what follows, when convenient, we will present our results for all three chemical systems, otherwise, we will focus upon SiC. We ultimately find analogous results across all chemistries, which suggests generalizability to any similar honeycomb bilayer system. Additional data for the BN and AlN systems can be found in~\cite{supp}.

\subsection{Global Symmetry Considerations}
\label{sec:a1}
Following the logic of~\cref{sec:f1}, we first lay out the macroscopic trends and symmetry details of our systems as they execute one period of sliding by using a combination of group theory and DFT calculations for a series of structures
across the period. These trends will then aid us in identifying our macroscopic parameter $\boldsymbol{\eta}$. The energy landscape across one period for each chemistry is depicted in~\cref{Fig:1a}, with structural diagrams for high symmetry states in SiC shown in~\cref{Fig:1b}. All chemistries share the same symmetry properties throughout the period: the $SP1$ structure is a 2D hexagonal lattice with $\mathcal{L}_{SP1}=p\bar{6}m2$ (no. 78) and PG $D_{3h}$. Its primitive unit cell contains one atom of each element per layer for a total of four atoms. Sliding can occur along three equivalent crystallographic directions, related to each other by the OOP threefold rotations of $D_{3h}$. For simplicity, we take the OOP direction along $\hat{z}$ and IP sliding to occur along [120], which corresponds to the Cartesian $\hat{y}$ direction. Therefore, one period of sliding corresponds to $s=\sqrt{3}a$ where $a$ is the lattice constant. 

As the system traverses a period, we encounter $\mathcal{L}_P=p3m1$ (no. 69) with PG $C_{3v}$, $\mathcal{L}_{SP2}=cem2$ (no. 36) with PG $C_{2v}$ and $\mathcal{L}_{INT}=c1m1$ (no. 13) with PG $C_s$. These symmetry changes are depicted diagrammatically in~\cref{Fig:1a}, with arrows denoting the existence of a group-subgroup relationship. As is the case in all previous reports of rigid bilayer sliding ferroelectricity~\cite{guo2024,yang2024}, the size of the primitive cell remains constant throughout the period.  

\begin{table}[]
	\centering
	\setlength{\tabcolsep}{4pt}
	\begin{tabular}{|c|c|c|}
		\hline
		Transition & Irrep & Allowed Partners \\
		\hline
		$SP1\rightarrow INT$ & $E^{''}(D_{3h})$ & $\{xz,yz\}$   \\
		\hline
		$SP2\rightarrow INT$ & $B_1(C_{2v})$ & $z$, $yz$  \\
		\hline
		$+P\rightarrow INT$ & $E(C_{3v})$ & $\{x,y\}$,  $\{x^2-y^2,xy\}$, $\{xz,yz\}$  \\
		\hline
	\end{tabular}
	\caption{For each transition, the corresponding irrep and polynomials, up to second order, which carry the irrep}
	\label{tab:oppolynomials}
\end{table}

Mentioned in~\cref{sec:f1}, the above symmetries alone can determine the polarization magnitude profiles $p_{\parallel}(s)=p_y(s)$ and $p_{\perp}(s)=p_z(s)$ as a function of the sliding distance, which are known to good approximation~\cite{bennett2023,supp} to be
\begin{align}
    & p_{\parallel}(s) \propto \cos{(ks)}-\cos{(2ks)} \label{eq:7} \\
    & p_{\perp}(s) \propto 2\sin{(ks)}-\sin{(2ks)} \label{eq:8} 
\end{align}
where $k=2\pi / \sqrt{3}a$. Our DFT-calculated profiles,~\cref{Fig:2a,Fig:2b}, show excellent agreement with these fits. 

Now, continually following our prescription in~\cref{sec:f1}, we use group theory to determine the irreps which induce the transitions $SP1\rightarrow INT$, $SP2\rightarrow INT$ and $+P\rightarrow INT$, and find the sets of functions which carry them. Then, to smoothly link these reconstructive transitions together, we enforce compatibility across the irreps by finding the partner functions that simultaneously carry all three; these functions and any others with the same transformation properties will act as $\boldsymbol{\eta}$. The results are displayed in~\cref{tab:oppolynomials}. Although these irreps can all be identified using group theory, we can confirm the symmetry behavior of the first two transitions, which correspond to sliding instabilities, by performing phonon calculations to find their driving soft modes. This is discussed in detail in Section SII. 

~\cref{tab:oppolynomials} shows that $yz$ belongs to the intersection of all three sets. Note that while the $E^{''}(D_{3h})$ and $E(C_{3v})$ irreps are two-dimensional, compatibility relations with the $SP2$ structure enforce that our overarching parameter of interest $\eta$ be one-dimensional, so we drop the bolding from our notation. Therefore, our global parameter is any function which transforms like $yz$ under all the symmetries encountered along the sliding period: 
\begin{equation}
	\label{eqn:OP}
	\eta \propto yz.
\end{equation}
Similar to $p_{\parallel}(s)$ and $p_{\perp}(s)$, the functional form of $\eta(s)$ can now be determined via the transformation properties of the polynomial $yz$ and is derived in Section SIV A~\cite{supp}:

\begin{equation}
	\label{eqn:opcurve}
	\eta(s) \propto \sin{(ks)}+\sin{(2ks)} .
\end{equation}

\begin{figure*}
	\includegraphics[width=\textwidth]{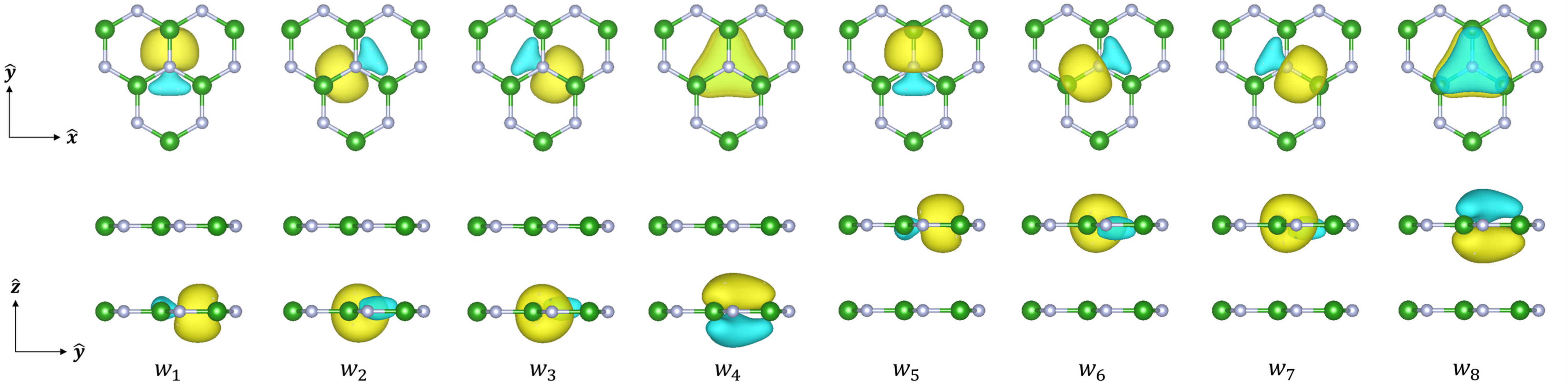}
	\caption{In-plane (IP) and out-of-plane (OOP) profiles (top and bottom, respectively) of example MLWFs $w_n$, $n\in\{1,2,...,8\}$, generated for the $SP1$ structure of SiC and plotted at the 0.1 \AA$^{-3/2}$ isosurface level. The MLWFs constructed for other structures along the sliding period all show these same general $sp^2$ and lone pair (LP) orbital characters. Yellow represents positive wavefunction and blue represents negative.}
    \label{Fig:3}
\end{figure*} 

Finally, studying invariant free energy polynomials $\mathcal{F}(\eta,p_{\parallel},p_{\perp})$ constructed with respect to the $SP1$ and $SP2$ states (see~\cref{tab:couplings}) reveals that the appearance of $p_{\perp}$ in the vicinity of a nonpolar structure can result from a variety of second and third-order couplings with $\eta$ and $p_{\parallel}$. This underscores the observation of previous studies~\cite{bennett2023} and the trends depicted in~\cref{Fig:2a,Fig:2b} that, in these systems, $p_{\parallel}$ and $p_{\perp}$ necessarily evolve simultaneously. These couplings can also provide additional insight into the polarization landscapes, for example, it makes sense in~\cref{Fig:2b} that $\frac{dp_{\perp}}{ds}$ is larger at $SP2$ than at $SP1$ since it results in part from a 2$^{nd}$-order coupling at $SP2$ and a 3$^{rd}$-order coupling at $SP1$.

\begin{table}[h]
	\centering
	\setlength{\tabcolsep}{4pt}
	\begin{tabular}{|c|c|c|}
		\hline
		Extremum & 2$^{nd}$ Order & 3$^{rd}$ Order \\
		\hline
		$SP1$  & n/a & $\eta^2 p_{\parallel},~\eta p_{\parallel}p_{\perp}$   \\
		\hline
		$SP2$  & $\eta p_{\perp}$ & $\eta^2 p_{\parallel},~p_{\parallel}p_{\perp}^2,~\eta p_{\parallel}p_{\perp}$  \\
		\hline
		$+P$ & $\eta p_{\parallel}$ & $\eta^2 p_{\parallel},~\eta^2 p_{\perp},~\eta p_{\parallel}^2,~p_{\parallel}^2p_{\perp},~\eta p_{\parallel}p_{\perp}$  \\
		\hline
	\end{tabular}
	\caption{Allowed nontrivial second and third order couplings in the polynomial $\mathcal{F}(\eta,p_{\parallel},p_{\perp})$ expanded about the $SP1$, $SP2$ and $+P$ extrema of the energy landscape.}
	\label{tab:couplings}
\end{table}	

\subsection{Evolution of MLWFs}
\label{sec:a2}
Now that we have a clear understanding of the symmetry properties that the polarization mechanism must obey, we follow~\cref{sec:f2} and construct MLWFs and DMDs as a starting point in determining the physical microscopic nature of the polarization mechanism. Monolayers of our honeycomb systems are known to exhibit (ignoring spin) three $sp^2$-like IP bonding orbitals and one OOP lone pair (LP) orbital, all of which are centered about the $A$ element~\cite{wang2023,sahin2009,ooi2005}. Therefore, for a bilayer structure, MLWF calculations result in eight orbitals $w_n(\mathbf{r})$ per unit cell, where we assign $n\in \{1,2,..,8\}$ as depicted in~\cref{Fig:3}. The bottom layer MLWFs, $n=1,2,3,4$, have top-layer counterparts $n=5,6,7,8$, respectively. This general trend of eight MLWFs per structure, with one LP and three $sp^2$-like per layer, applies to all stacking configurations sampled within a period because the stacking perturbation is too weak to significantly alter the overriding character of the orbitals.

 To confirm our MLWFs are reasonable, we use the modern theory of polarization to compute the electronic polarization profiles $p_{\parallel}(s)$ and $p_{\perp}(s)$ using Wannier centers~\cite{kingsmith_1993},
\begin{equation}
\mathbf{P_{el}}=\frac{-e}{V_\text{cell}}\sum_n \mathbf{r_W}_n, 
\label{eqn:wannier_pol}
\end{equation}
where $V_\text{cell}$ is the unit cell volume and $\mathbf{r_W}_n$ is the $n^{th}$ Wannier center. Depicted in~\cref{Fig:2a,Fig:2b} with dashed lines, they show good agreement with those computed using DFT. Finally, we use our MLWFs to calculate DMDs using~\cref{eqn:dmlwfd}.

To pare down these eight MLWFs into the set $W$ of relevant DMDs, we intuitively expect that the LP MLWFs $w_4$ and $w_8$ should be most pertinent to the bilayer sliding problem since their spatial extent into the interlayer region is much greater than that of the $sp^2$-like orbitals. We confirm this by calculating $\langle z^2 \rangle$ for all chemistries and orbitals in Section SI~\cite{supp}. This agrees with several previous reports that have attributed sliding ferroelectricity to action by the OOP orbitals~\cite{wang2023,yasuda2021,zhi2025,m.liu2023}. LP orbitals are also known to play an important role in the ferroelectric transition of some bulk compounds~\cite{smith2015,shen2019}.

\begin{figure*}
\centering

    \includegraphics[scale=0.265]{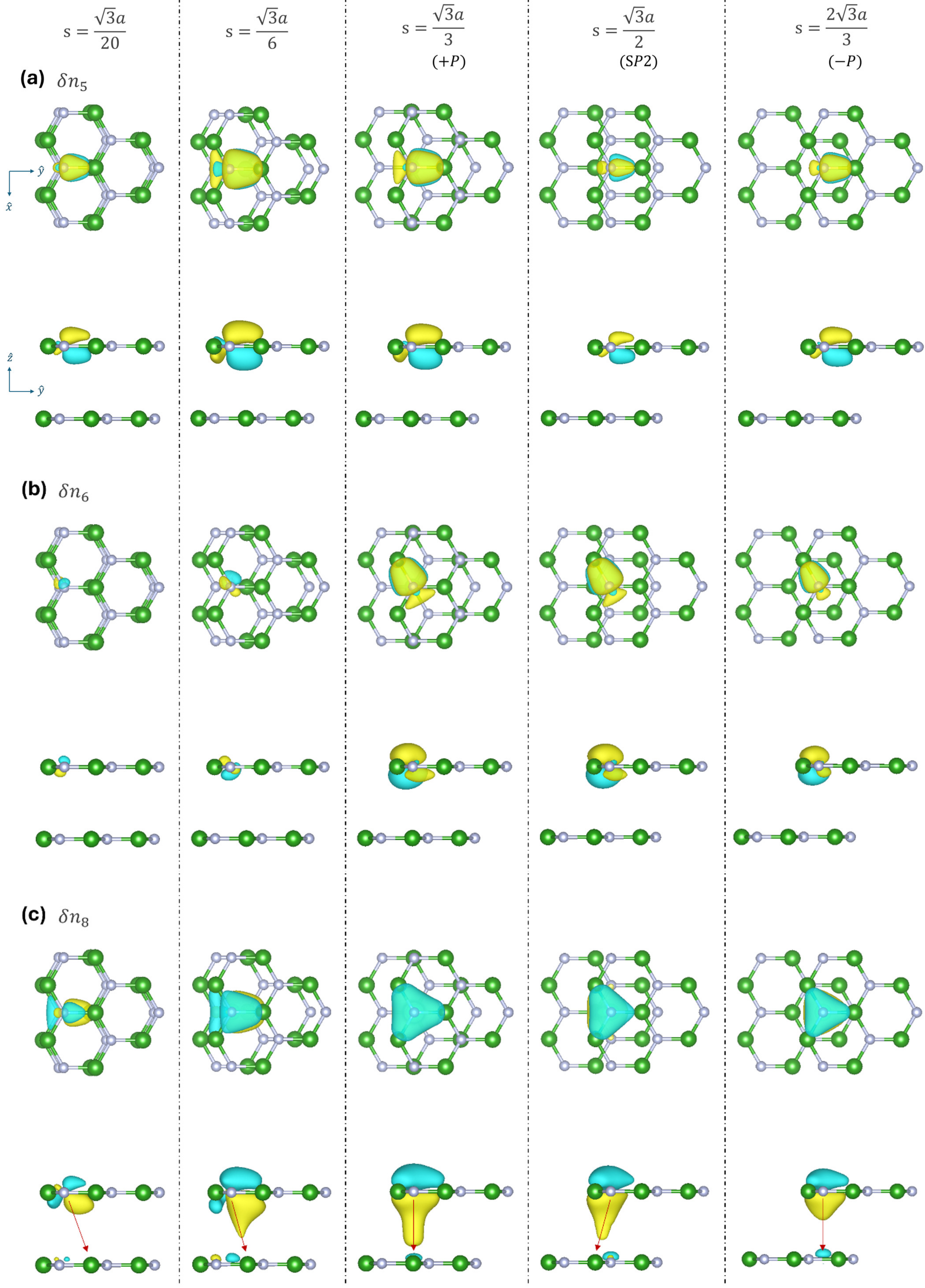}
    \caption{IP and OOP views of the differential MLWF densities (DMDs) a) $\delta n_5(s)$ b) $\delta n_6(s)$ and c) $\delta n_8(s)$ for a selection of five $s$ along a sliding period in SiC. Yellow (blue) regions represent areas of electron density accumulation (depletion) with respect to the reference densities. All plots are taken at the 6.2$\times 10^{-4}$ \AA$^{-3}$ isosurface level. Comparing a,b,c) shows that the density distortions of $w_8(\mathbf{r})$ behave differently than those of $w_5(\mathbf{r})$ and $w_6(\mathbf{r})$, exhibiting a greater extent in the interlayer region with some directionality toward the opposing layer's $B$ atom (green), indicated by the thin arrows.}
    \label{Fig:4} 
\end{figure*}

To bolster this intuition with the  energetic analysis described in~\cref{sec:f2}, calculations for SiC, AlN and BN show that the minimum of a monolayer potential is located at the $B$ atom (see Fig. S2~\cite{supp}). Therefore, following~\cref{sec:f2}, if a MLWF density's interlayer interactions dominate, its distortion should largely be directed toward the opposite layer's $B$ atom. In this way, the DMD plots in~\cref{Fig:4} further suggest that LP orbitals are the important MLWFs since their density accumulations appear to be directed toward the opposing layer's $B$ atom (Si, in green), while those of the $sp^2$-like orbitals show no such trend. In fact, the $sp^2$ orbital densities seem to exhibit secondary distortions, primarily driven by intralayer repulsive interactions with the LP densities since regions of accumulation (depletion) in the $sp^2$ DMDs mostly correspond to regions of depletion (accumulation) in the LP DMDs. Similar trends persist for AlN and BN, as shown in Section SV~\cite{supp}.

We can roughly quantify these directional trends for each DMD by gathering the set of points in space showing the largest electron density increases and performing truncated singular value decomposition upon it to determine the principal unit vector direction of significant density accumulation. We expect this vector, which we call $\mathbf{d}$, to point roughly toward the closest $B$ atom in the opposing layer if that orbital's distortion is driven primarily by interlayer interactions. Namely,  $\mathbf{d}\cdot\mathbf{t}\approx1$ for relevant MLWFs, where $\mathbf{t}$ is the unit vector pointing from the origin of the $\mathbf{r_{I_i}}$ coordinate system to the $B$ atom in the opposite layer. ~\cref{tab:3} shows this dot product calculated for $\delta n_5$, $\delta n_6$ and $\delta n_8$ in SiC for a few $s$ from~\cref{Fig:4}. For $\delta n_8$, it is much closer to 1 across $s$ than it is for the other MLWFs, suggesting again that LP electrons are the most consistent with distortions driven primarily by interlayer rather than intralayer energetics. 

\begin{table}[h]
	\centering
	\setlength{\tabcolsep}{4pt}
	\begin{tabular}{|c|c|c|c|c|}
		\hline
		MLWF & $s=\sqrt{3}a/20$ & $s=\sqrt{3}a/6$ & $s=\sqrt{3}a/3$ & $s=\sqrt{3}a/2$ \\
		\hline
		5 & -0.372 & -0.484 & -0.641 & -0.746 \\
		\hline
		6 & 0.295 & -0.629 & -0.634 & -0.554\\
		\hline
		8 & 0.920 & 0.986 & 1.000 & 1.000 \\
		\hline
		
	\end{tabular}
	\caption{$\mathbf{d}(s)\cdot\mathbf{t}(s)$ for $\delta n_i (s),~i \in \{5,6,8\}$ in SiC.} 
	\label{tab:3}
\end{table}

All the above suggests that we treat the LP orbital densities of MLWFs $w_4$ and $w_8$ as the most relevant local electronic components in determining the origin of the electronic response to sliding:
\begin{equation}
	\label{eqn:w}
	W=\{\delta n_4,~\delta n_8\}.
\end{equation}

\subsection{Connecting the Local and Global Pictures}
\label{sec:a3}
With our prominent DMD contributions identified, we continue in the vein of~\cref{sec:f3} to link them with the macroscopic parameter $\eta$ by first performing a multipole decomposition and then relating the ensuing local symmetries to global bilayer-referenced symmetries. This will ultimately result in a physically meaningful representation of $\eta$ in terms of electron density distortions. We perform the spherical harmonic decomposition given by~\cref{eqn:ylm_decomp} on $\delta n_4$ and $\delta n_8$, using the position of the $A$ atom in each layer as the origin of each spherical coordinate system since the DMDs are largest in magnitude in the immediate vicinity of the $A$ atoms (see Fig. S1~\cite{supp}). 

\begin{figure*}[ht]
	\centering
	\subfloat{%
		\begin{minipage}{.97\textwidth}
			\includegraphics[width=\textwidth]{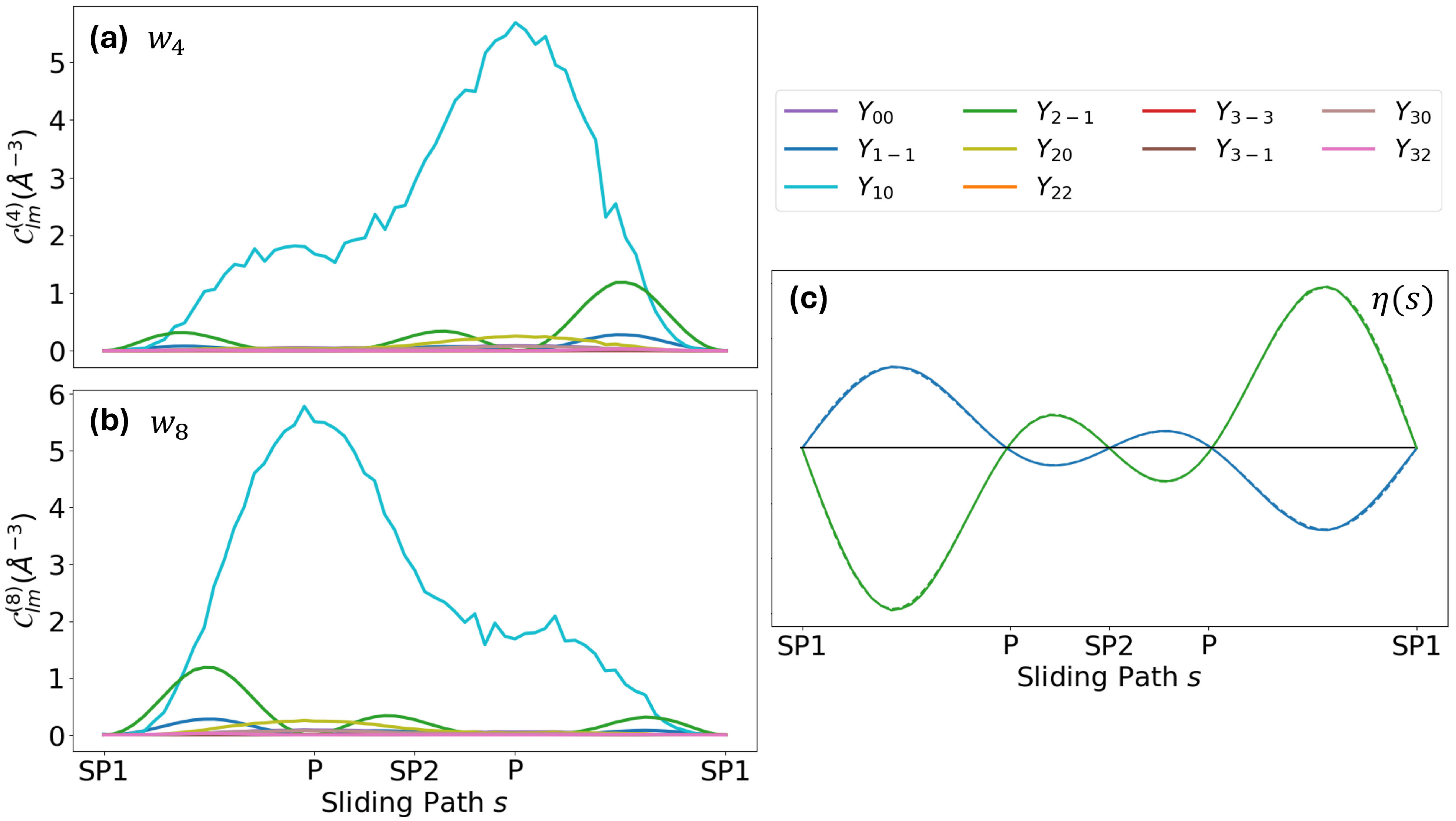}
		\end{minipage}%
	\label{Fig:finala}}
	\subfloat{%
		\begin{minipage}{0\textwidth}
			\includegraphics[width=\textwidth]{final-eps-converted-to.pdf}
		\end{minipage}%
	\label{Fig:finalb}}
	\subfloat{%
		\begin{minipage}{0\textwidth}
			\includegraphics[width=\textwidth]{final-eps-converted-to.pdf}
		\end{minipage}%
	\label{Fig:finalc}}
	\caption{$\mathcal{C}^{(i)}_{lm}(s)$ for up to $l=3$ terms in~\cref{eqn:ylm_decomp} for a) $\delta n_4$ and b) $\delta n_8$ in SiC. c) The top and bottom lines of~\cref{eqn:finaleta} plotted with blue and green solid lines, respectively, in arbitrary units. Dashed lines represent fits to the expected behavior of~\cref{eqn:opcurve}.}
\end{figure*}

To reduce the number of components we need to analyze, following~\cref{sec:f3}, we calculate and plot $\mathcal{C}^{(4,8)}_{lm}(s)$ across several $l,m$, shown in~\cref{Fig:finala,Fig:finalb} for SiC and Section SV~\cite{supp} for AlN and BN. For all chemistries, only the $Y_{1,0},~Y_{1,-1},~Y_{2,0}$ and $Y_{2,-1}$ components ever make nonnegligible contributions to $\delta n_4$ and $\delta n_8$ across $s$, so in the notation of~\cref{sec:f3},  $Y^{(4,8)}=\{Y_{10}^{(4,8)},~Y_{1-1}^{(4,8)},~Y_{20}^{(4,8)},~Y_{2-1}^{(4,8)}\}$. This gives us a basis $B$ of principal microscopic distortion components from which we construct a physical $\eta$:
\begin{equation}
	\label{eqn:b}
	B=\{Y_{10}^{(4)},~Y_{1-1}^{(4)},~Y_{20}^{(4)},~Y_{2-1}^{(4)},Y_{10}^{(8)},~Y_{1-1}^{(8)},~Y_{20}^{(8)},~Y_{2-1}^{(8)}\}.
\end{equation}

Then, we use $B$ as the basis for a rep of each PG encountered as we slide: $SP1$, $+P$, $SP2$ and $INT$. For the $C_{2v}$ PG of $SP2$ and the $C_s$ PG of $INT$, this basis is valid. For those of the $SP1$ and $\pm P$ structures, $Y_{1-1}$ and $Y_{2-1}$ are one of a pair of partners carrying 2D irreps, with the others being $Y_{11}$ and $Y_{21}$, respectively. Due to the coordinate system choice that sliding is along $\hat{y}$, these components are never allowed and thus do not show up in~\cref{eqn:ylm_decomp}, but need to be considered for the sake of creating a rep basis. Therefore, the set \{$Y_{11}^{(4)},~Y_{21}^{(4)},~Y_{11}^{(8)},~Y_{21}^{(8)}\}$ is added to $B$. 

The character tables for the resulting 12-dimensional reps of the PGs are displayed in Fig. S5~\cite{supp}, along with their irrep decompositions. All the irreps determined in~\cref{sec:a1} to be responsible for describing symmetry changes along the sliding period (see~\cref{tab:oppolynomials}) are present in the decompositions: $E^{''}(D_{3h})$, $E(C_{3v})$ and $B_1(C_{2v})$. Also, trivially, the symmetry breaking of sliding transforms like the identity irrep $A'$ of $INT$. Therefore, to find our global parameter, we construct a partner basis for these four irreps by finding which functions or linear combinations of functions in $B$ carry each of them.

Similar to~\cref{sec:a1}, the subset of functions which are shared among the four sets of partner functions will transform like the abstract $\eta$ of~\cref{eqn:OP}. These will be the symmetry-adapted combinations of relevant local distortions which characterize the macroscopic symmetry changes of sliding. By applying the projection operators and enforcing compatibility along the sliding pathway, we get the final result 
\begin{align}
	\label{eqn:finalresult}
	\eta \propto~ & Y_{1-1}^{(8)}-Y_{1-1}^{(4)}, \\
	&  Y_{2-1}^{(8)}+Y_{2-1}^{(4)}. \nonumber
\end{align}

So, our approach leads to the novel conclusion that sliding ferroelectricity in honeycomb bilayers is triggered by both anti-aligned dipole-like distortions and aligned quadrupole-like distortions of LP orbitals. To express how these distortions vary with $s$, we form the functions $\eta^{(4)}_{1-1}(s),~\eta^{(4)}_{2-1}(s),~\eta^{(8)}_{1-1}(s)$ and $\eta^{(8)}_{2-1}(s)$ via~\cref{eqn:eta} and take linear combinations in accordance with~\cref{eqn:finalresult}: 
\begin{align}
	\eta (s) \propto~ & \eta_{1-1}^{(8)}(s)-\eta_{1-1}^{(4)}(s), 	\label{eqn:finaleta} \\
	&  \eta_{2-1}^{(8)}(s)+\eta_{2-1}^{(4)}(s). \nonumber
\end{align}
As shown in \cref{Fig:finalc} for SiC and Section SV~\cite{supp} for AlN and BN, plotting these quantities yields two profiles which both fit extremely well to the functional form~\cref{eqn:opcurve} calculated using only symmetry considerations. Finally, to generate a $p_{\perp}$, these distortions couple to objects behaving like $z$-polar vectors through the couplings given in~\cref{tab:couplings}. 

\section{Discussion}
%%%not edited yet
It's interesting that LP orbitals, which in \cref{Fig:3} resemble C $p_z$ orbitals with slight contributions from Si $p_z$ orbitals, can become distorted in the $\hat{y}$ direction in addition to $\hat{z}$ to create the $Y_{1-1}$ and $Y_{2-1}$ contributions in~\cref{eqn:finalresult}. One simple explanation would be hybridization between the LP orbitals in one layer and nearby orbitals in the opposite layer as sliding proceeds. However, as depicted in Fig. S8~\cite{supp} for SiC and in agreement with previous work~\cite{m.liu2023}, our LP MLWFs show no evidence of hybridizing with any wavefunctions centered about the opposite layer. Instead, hybridization must occur between the LP and $sp^2$-like orbitals within the same layer, induced by the action of the opposing layer. As explained in Section SIII, the intralayer hybridization process and generation of a distortion in $\hat{y}$ can be understood by appealing to crystal field splitting.

The form of $\eta$ in~\cref{eqn:OP,eqn:finalresult} makes it clear that the traditional method of polarization switching in sliding ferroelectrics, by applying an OOP electric field, is inefficient because the distortions which cause the transition don't transform like $z$. This is also evident from laser excitation experiments~\cite{yang2024} which achieve 1000$\times$ faster polarization switching times than in other sliding ferroelectric studies~\cite{yasuda2024} because they couple to the relevant symmetry modes. In this way, sliding ferroelectrics are akin to improper ferroelectrics, which explains, for example, their robustness against depolarization fields~\cite{sai2009}. The form of~\cref{eqn:finalresult} suggests that either the application of a uniform quadrupolar field or opposing vector fields (for example, opposing IP strain gradients along $\mathbf{\hat{y}}$) would achieve more efficient switching. However,~\cref{Fig:finala,Fig:finalb,Fig:finalc} make clear that the quadrupolar distortions are stronger than the dipolar ones, suggesting that preferential focus should be placed upon inducing quadrupolar-like distortions in each layer to achieve the most efficient switching.

In all, we have presented a rigorous approach involving group theory and DFT to identify the primary, symmetry-adapted electronic mechanism which causes the emergence of a polarization in sliding ferroelectrics. We have applied our strategy to honeycomb bilayer systems to find that LP orbitals, and their tendency toward developing IP, antialigned vector-like distortions and aligned quadrupole-like distortions, induce the polarization. As honeycomb bilayers are the simplest sliding ferroelectrics, this approach represents a foundation which can be built upon to generalize to more complex systems, for example, those involving more complex formula units, higher occupied orbital states and more than two layers. As the fundamental mechanisms behind sliding ferroelectricity in these systems become better elucidated, materials designers will be better equipped to leverage them to achieve desired functional and performance capabilities.

\section{Computational Details}

DFT calculations were performed using the ABINIT~\cite{gonze2020} suite, with the exchange-correlation functional treated under the revised generalized gradient approximation of Perdew, Burke and Ernzerhof (PBE)~\cite{perdew1996}.  AlN, BN and SiC bilayers were modeled as slabs separated by large ($\approx$ 24 \AA) vacuum regions, with vdW interactions included via the DFT-D3 method~\cite{vantroeye2016}. All calculations used a total energy difference threshold of 2.7$\times10^{-9}$ eV between cycles, achieved twice consecutively, for the self-consistent stopping criterion and a 1.9 keV plane wave cutoff. Reciprocal space was sampled using a 20$\times$20$\times$1 $\Gamma$-centered mesh for AlN and SiC, and a 24$\times$24$\times$1  $\Gamma$-centered mesh for BN due to its smaller IP lattice constant. To find equilibrium IP lattice constants and internal coordinates, structures were relaxed until the residual Hellman-Feynman on each atom was less than 2.6$\times10^{-4}$ eV/\AA. Sliding periods were sampled via a series of 62 structures separated by increments of $s=\sqrt{3}a/62$.

Within ABINIT, phonon calculations were performed using density function perturbation theory (DFPT), and polarizations were calculated using the Berry phase method. MLWFs were constructed using ABINIT’s interface with the wannier90 program~\cite{mostofi2014} by first making an initial projection onto $sp^2$ and $p_z$ orbitals centered on the $A$ atoms and then converging until the variation in the total spread of the wavefunctions fell below 10$\times10^{-10}$~\AA$^2$ over three successive iterations.

\section{Acknowledgements}

We acknowledge support from the National Science Foundation Platform for the Accelerated Realization, Analysis, and Discovery of Interface Materials (PARADIM) under Cooperative Agreement No. DMR-2039380. Significant credit for this work should also be directed toward the late Craig J. Fennie, who guided this project from its inception to its near completion.

\bibliography{references}

@PREAMBLE{
 "\providecommand{\noopsort}[1]{}" 
 # "\providecommand{\singleletter}[1]{#1}%" 
}

@article{li2017,
  title={Binary compound bilayer and multilayer with vertical polarizations: two-dimensional ferroelectrics, multiferroics, and nanogenerators},
  author={Li, Lei and Wu, Menghao},
  journal={ACS nano},
  volume={11},
  number={6},
  pages={6382--6388},
  year={2017},
  publisher={ACS Publications},
  doi={10.1021/acsnano.7b02756}}

@article{wang2023,
  title={Sliding ferroelectricity in bilayer honeycomb structures: A first-principles study},
  author={Wang, Zhe and Gui, Zhigang and Huang, Li},
  journal={Physical Review B},
  volume={107},
  number={3},
  pages={035426},
  year={2023},
  publisher={APS},
  doi={10.1103/PhysRevB.107.035426}}

@article{wu2021,
  title={Sliding ferroelectricity in 2D van der Waals materials: Related physics and future opportunities},
  author={Wu, Menghao and Li, Ju},
  journal={Proceedings of the National Academy of Sciences},
  volume={118},
  number={50},
  pages={e2115703118},
  year={2021},
  publisher={National Acad Sciences},
  doi={10.1073/pnas.2115703118}}

@article{zhang2023,
  title={Ferroelectric order in van der Waals layered materials},
  author={Zhang, Dawei and Schoenherr, Peggy and Sharma, Pankaj and Seidel, Jan},
  journal={Nature Reviews Materials},
  volume={8},
  number={1},
  pages={25--40},
  year={2023},
  publisher={Nature Publishing Group UK London},
  doi={10.1038/s41578-022-00484-3}}

@article{s.li2024,
  title={Van der Waals Ferroelectrics: Theories, Materials, and Device Applications},
  author={Li, Shuhui and Wang, Feng and Wang, Yanrong and Yang, Jia and Wang, Xinyuan and Zhan, Xueying and He, Jun and Wang, Zhenxing},
  journal={Advanced Materials},
  volume={36},
  number={22},
  pages={2301472},
  year={2024},
  publisher={Wiley Online Library},
  doi={10.1002/adma.202301472}}

@article{bian2023,
  title={High-performance sliding ferroelectric transistor based on schottky barrier tuning},
  author={Bian, Renji and Cao, Guiming and Pan, Er and Liu, Qing and Li, Zefen and Liang, Lei and Wu, Qingyun and Ang, Lay Kee and Li, Wenwu and Zhao, Xiaoxu and others},
  journal={Nano Letters},
  volume={23},
  number={10},
  pages={4595--4601},
  year={2023},
  publisher={ACS Publications},
  doi={10.1021/acs.nanolett.3c01053}}

@article{yasuda2024,
  title={Ultrafast high-endurance memory based on sliding ferroelectrics},
  author={Yasuda, Kenji and Zalys-Geller, Evan and Wang, Xirui and Bennett, Daniel and Cheema, Suraj S and Watanabe, Kenji and Taniguchi, Takashi and Kaxiras, Efthimios and Jarillo-Herrero, Pablo and Ashoori, Raymond},
  journal={Science},
  volume={0},
  pages={eadp3575},
  year={2024},
  publisher={American Association for the Advancement of Science},
  doi={10.1126/science.adp3575}}

@article{x.li2024,
	author={Li, Xiuzhen and Qin, Biao and Wang, Yaxian and Xi, Yue and Huang, Zhiheng and Zhao, Mengze and Peng, Yalin and Chen, Zitao and Pan, Zitian and Zhu, Jundong and others},
	journal={Nature Communications},
	volume={15},
	number={1},
	pages={10921},
	year={2024},
	publisher={Nature Publishing Group UK London},
	doi={10.1038/s41467-024-55333-4}}

@article{jin2022,
  title={Ferroelectrics-integrated two-dimensional devices toward next-generation electronics},
  author={Jin, Tengyu and Mao, Jingyu and Gao, Jing and Han, Cheng and Loh, Kian Ping and Wee, Andrew TS and Chen, Wei},
  journal={ACS nano},
  volume={16},
  number={9},
  pages={13595--13611},
  year={2022},
  publisher={ACS Publications},
  doi={10.1021/acsnano.2c07281}}

@article{dai2019,
  title={Robust piezo-phototronic effect in multilayer $\gamma$-InSe for high-performance self-powered flexible photodetectors},
  author={Dai, Mingjin and Chen, Hongyu and Wang, Fakun and Hu, Yunxia and Wei, Shuai and Zhang, Jia and Wang, Zhiguo and Zhai, Tianyou and Hu, PingAn},
  journal={ACS nano},
  volume={13},
  number={6},
  pages={7291--7299},
  year={2019},
  publisher={ACS Publications},
  doi={10.1021/acsnano.9b03278}}

@article{sun2022,
  title={Mesoscopic sliding ferroelectricity enabled photovoltaic random access memory for material-level artificial vision system},
  author={Sun, Yan and Xu, Shuting and Xu, Zheqi and Tian, Jiamin and Bai, Mengmeng and Qi, Zhiying and Niu, Yue and Aung, Hein Htet and Xiong, Xiaolu and Han, Junfeng and others},
  journal={Nature communications},
  volume={13},
  number={1},
  pages={5391},
  year={2022},
  publisher={Nature Publishing Group UK London},
  doi={10.1038/s41467-022-33118-x}}

@article{xiao2022,
  title={Non-synchronous bulk photovoltaic effect in two-dimensional interlayer-sliding ferroelectrics},
  author={Xiao, Rui-Chun and Gao, Yang and Jiang, Hua and Gan, Wei and Zhang, Changjin and Li, Hui},
  journal={npj Computational Materials},
  volume={8},
  number={1},
  pages={138},
  year={2022},
  publisher={Nature Publishing Group UK London},
  doi={10.1038/s41524-022-00828-1}}

@article{xu2022,
  title={Van der Waals force-induced intralayer ferroelectric-to-antiferroelectric transition via interlayer sliding in bilayer group-IV monochalcogenides},
  author={Xu, Bo and Deng, Junkai and Ding, Xiangdong and Sun, Jun and Liu, Jefferson Zhe},
  journal={npj Computational Materials},
  volume={8},
  number={1},
  pages={47},
  year={2022},
  publisher={Nature Publishing Group UK London},
  doi={10.1038/s41524-022-00724-8}}

@article{qi2021,
  title={Review on recent developments in 2D ferroelectrics: Theories and applications},
  author={Qi, Lu and Ruan, Shuangchen and Zeng, Yu-Jia},
  journal={Advanced Materials},
  volume={33},
  number={13},
  pages={2005098},
  year={2021},
  publisher={Wiley Online Library},
  doi={10.1002/adma.202005098}}

@article{jia2022,
  title={Flexible ferroelectric devices: status and applications},
  author={Jia, Xiaotong and Guo, Rui and Tay, Beng Kang and Yan, Xiaobing},
  journal={Advanced Functional Materials},
  volume={32},
  number={45},
  pages={2205933},
  year={2022},
  publisher={Wiley Online Library},
  doi={10.1002/adfm.202205933}}

@article{fei2018,
  title={Ferroelectric switching of a two-dimensional metal},
  author={Fei, Zaiyao and Zhao, Wenjin and Palomaki, Tauno A and Sun, Bosong and Miller, Moira K and Zhao, Zhiying and Yan, Jiaqiang and Xu, Xiaodong and Cobden, David H},
  journal={Nature},
  volume={560},
  number={7718},
  pages={336--339},
  year={2018},
  publisher={Nature Publishing Group UK London},
  doi={10.1038/s41586-018-0336-3}}

@article{yang2018,
  title={Origin of two-dimensional vertical ferroelectricity in WTe2 bilayer and multilayer},
  author={Yang, Qing and Wu, Menghao and Li, Ju},
  journal={The journal of physical chemistry letters},
  volume={9},
  number={24},
  pages={7160--7164},
  year={2018},
  publisher={ACS Publications},
  doi={10.1021/acs.jpclett.8b03654}}

@article{jindal2023,
  title={Coupled ferroelectricity and superconductivity in bilayer Td-MoTe2},
  author={Jindal, Apoorv and Saha, Amartyajyoti and Li, Zizhong and Taniguchi, Takashi and Watanabe, Kenji and Hone, James C and Birol, Turan and Fernandes, Rafael M and Dean, Cory R and Pasupathy, Abhay N and others},
  journal={Nature},
  volume={613},
  number={7942},
  pages={48--52},
  year={2023},
  publisher={Nature Publishing Group UK London},
  doi={10.1038/s41586-022-05521-3}}

@article{yasuda2021,
  title={Stacking-engineered ferroelectricity in bilayer boron nitride},
  author={Yasuda, Kenji and Wang, Xirui and Watanabe, Kenji and Taniguchi, Takashi and Jarillo-Herrero, Pablo},
  journal={Science},
  volume={372},
  number={6549},
  pages={1458--1462},
  year={2021},
  publisher={American Association for the Advancement of Science},
  doi={10.1126/science.abd3230}}

@article{viznerstern2021,
  title={Interfacial ferroelectricity by van der Waals sliding},
  author={Vizner Stern, Maayan and Waschitz, Yuval and Cao, Wei and Nevo, Iftach and Watanabe, Kenji and Taniguchi, Takashi and Sela, Eran and Urbakh, Michael and Hod, Oded and Ben Shalom, Moshe},
  journal={Science},
  volume={372},
  number={6549},
  pages={1462--1466},
  year={2021},
  publisher={American Association for the Advancement of Science},
  doi={10.1126/science.abe8177}}

@article{k.liu2023,
  title={Tunable sliding ferroelectricity and magnetoelectric coupling in two-dimensional multiferroic MnSe materials},
  author={Liu, Kehan and Ma, Xikui and Xu, Shuoke and Li, Yangyang and Zhao, Mingwen},
  journal={npj Computational Materials},
  volume={9},
  number={1},
  pages={16},
  year={2023},
  publisher={Nature Publishing Group UK London},
  doi={10.1038/s41524-023-00972-2}}

@article{yu2024,
  title={Electrical control of noncollinear magnetism in VAl2S4 van der Waals structures},
  author={Yu, Shiqiang and Xu, Yushuo and Dai, Ying and Sun, Dongyue and Huang, Baibiao and Wei, Wei},
  journal={Applied Physics Letters},
  volume={124},
  number={21},
  year={2024},
  pages={212903},
  publisher={AIP Publishing},
  doi={10.1063/5.0195872}}

@article{meng2022,
  title={Sliding induced multiple polarization states in two-dimensional ferroelectrics},
  author={Meng, Peng and Wu, Yaze and Bian, Renji and Pan, Er and Dong, Biao and Zhao, Xiaoxu and Chen, Jiangang and Wu, Lishu and Sun, Yuqi and Fu, Qundong and others},
  journal={Nature Communications},
  volume={13},
  number={1},
  pages={7696},
  year={2022},
  publisher={Nature Publishing Group UK London},
  doi={10.1038/s41467-022-35339-6}}

@article{wang2022,
  title={Interfacial ferroelectricity in rhombohedral-stacked bilayer transition metal dichalcogenides},
  author={Wang, Xirui and Yasuda, Kenji and Zhang, Yang and Liu, Song and Watanabe, Kenji and Taniguchi, Takashi and Hone, James and Fu, Liang and Jarillo-Herrero, Pablo},
  journal={Nature nanotechnology},
  volume={17},
  number={4},
  pages={367--371},
  year={2022},
  publisher={Nature Publishing Group UK London},
  doi={10.1038/s41565-021-01059-z}}

@article{wan2022,
  title={Room-Temperature Ferroelectricity in 1 T′-ReS 2 Multilayers},
  author={Wan, Yi and Hu, Ting and Mao, Xiaoyu and Fu, Jun and Yuan, Kai and Song, Yu and Gan, Xuetao and Xu, Xiaolong and Xue, Mingzhu and Cheng, Xing and others},
  journal={Physical Review Letters},
  volume={128},
  number={6},
  pages={067601},
  year={2022},
  publisher={APS},
  doi={10.1103/PhysRevLett.128.067601}}

@article{li2023,
  title={Room-temperature vertical ferroelectricity in rhenium diselenide induced by interlayer sliding},
  author={Li, Fang and Fu, Jun and Xue, Mingzhu and Li, You and Zeng, Hualing and Kan, Erjun and Hu, Ting and Wan, Yi},
  journal={Frontiers of Physics},
  volume={18},
  number={5},
  pages={53305},
  year={2023},
  publisher={Springer},
  doi={10.1007/s11467-023-1304-4}}

@article{sharma2019,
  title={A room-temperature ferroelectric semimetal},
  author={Sharma, Pankaj and Xiang, Fei-Xiang and Shao, Ding-Fu and Zhang, Dawei and Tsymbal, Evgeny Y and Hamilton, Alex R and Seidel, Jan},
  journal={Science advances},
  volume={5},
  number={7},
  pages={eaax5080},
  year={2019},
  publisher={American Association for the Advancement of Science},
  doi={10.1126/sciadv.aax5080}}

@article{ran2024,
  title={Moir{\'e} Ferroelectricity in Twisted Multilayer SnSe2},
  author={Ran, Yutong and Meng, Chen and Lu, Ziao and Wang, Huaipeng and Ma, Yunpeng and Xie, Dan and Li, Qian and Zhu, Hongwei},
  journal={Small Structures},
  pages={2400621},
  year={2024},
  publisher={Wiley Online Library},
  doi={10.1002/sstr.202400621}}

@article{liu2019,
  title={Vertical ferroelectric switching by in-plane sliding of two-dimensional bilayer WTe2},
  author={Liu, Xingen and Yang, Yali and Hu, Tao and Zhao, Guodong and Chen, Chen and Ren, Wei},
  journal={Nanoscale},
  volume={11},
  number={40},
  pages={18575--18581},
  year={2019},
  publisher={The Royal Society of Chemistry},
  doi={10.1039/C9NR05404A}}

@article{lin2024,
  title={Sliding ferroelectricity and the moir{\'e} effect in Janus bilayer MoSSe},
  author={Lin, Liyan and Hu, Xueqin and Meng, Ruijie and Li, Xu and Guo, Yandong and Da, Haixia and Jiang, Yue and Wang, Dongdong and Yang, Yurong and Yan, Xiaohong},
  journal={Nanoscale},
  volume={16},
  number={9},
  pages={4841--4850},
  year={2024},
  publisher={Royal Society of Chemistry},
  doi={10.1039/D3NR05730E}}

@article{rogee2022,
  title={Ferroelectricity in untwisted heterobilayers of transition metal dichalcogenides},
  author={Rog{\'e}e, Lukas and Wang, Lvjin and Zhang, Yi and Cai, Songhua and Wang, Peng and Chhowalla, Manish and Ji, Wei and Lau, Shu Ping},
  journal={Science},
  volume={376},
  number={6596},
  pages={973--978},
  year={2022},
  publisher={American Association for the Advancement of Science},
  doi={10.1126/science.abm5734}}

@article{m.liu2023,
  title={Orbital distortion and electric field control of sliding ferroelectricity in a boron nitride bilayer},
  author={Liu, Meng and Ji, Hongyan and Fu, Zhaoming and Wang, Yeliang and Sun, Jia-Tao and Gao, Hong-Jun},
  journal={Journal of Physics: Condensed Matter},
  volume={35},
  number={23},
  pages={235001},
  year={2023},
  publisher={IOP Publishing},
  doi={10.1088/1361-648X/acc561}}

@article{ma2021,
  title={Tunable vertical ferroelectricity and domain walls by interlayer sliding in $\beta$-ZrI2},
  author={Ma, Xiaonan and Liu, Chang and Ren, Wei and Nikolaev, Sergey A},
  journal={npj Computational Materials},
  volume={7},
  number={1},
  pages={177},
  year={2021},
  publisher={Nature Publishing Group UK London},
  doi={10.1038/s41524-021-00648-9}}

@article{zhong2021,
  title={Sliding ferroelectricity in two-dimensional MoA 2 N 4 (A= Si or Ge) bilayers: high polarizations and Moir{\'e} potentials},
  author={Zhong, Tingting and Ren, Yangyang and Zhang, Zhuhua and Gao, Jinhua and Wu, Menghao},
  journal={Journal of Materials Chemistry A},
  volume={9},
  number={35},
  pages={19659--19663},
  year={2021},
  publisher={Royal Society of Chemistry},
  doi={10.1039/D1TA02645C}}

@article{li2022,
  title={Enhanced vertical polarization and ultra-low polarization switching barriers of two-dimensional SnS/SnSSe ferroelectric heterostructures},
  author={Li, Yun-Qin and Wang, Xin-Yu and Zhu, Shi-Yu and Tang, Dai-Song and He, Qi-Wen and Wang, Xiao-Chun},
  journal={Journal of Materials Chemistry C},
  volume={10},
  number={33},
  pages={12132--12140},
  year={2022},
  publisher={Royal Society of Chemistry},
  doi={10.1039/D2TC02721F}}

@article{yang2023a,
  title = {Atypical Sliding and Moir\'e Ferroelectricity in Pure Multilayer Graphene},
  author = {Yang, Liu and Ding, Shiping and Gao, Jinhua and Wu, Menghao},
  journal = {Phys. Rev. Lett.},
  volume = {131},
  issue = {9},
  pages = {096801},
  numpages = {7},
  year = {2023},
  month = {Aug},
  publisher = {American Physical Society},
  doi = {10.1103/PhysRevLett.131.096801}}

@article{yang2023b,
  title={Across-layer sliding ferroelectricity in 2D heterolayers},
  author={Yang, Liu and Wu, Menghao},
  journal={Advanced Functional Materials},
  volume={33},
  number={29},
  pages={2301105},
  year={2023},
  publisher={Wiley Online Library},
  doi={10.1002/adfm.202301105}}

@article{freeman2006,
  title = {Graphitic Nanofilms as Precursors to Wurtzite Films: Theory},
  author = {Freeman, Colin L. and Claeyssens, Frederik and Allan, Neil L. and Harding, John H.},
  journal = {Phys. Rev. Lett.},
  volume = {96},
  issue = {6},
  pages = {066102},
  numpages = {4},
  year = {2006},
  month = {Feb},
  publisher = {American Physical Society},
  doi = {10.1103/PhysRevLett.96.066102}}

@article{lin2012,
  title={Light-emitting two-dimensional ultrathin silicon carbide},
  author={Lin, SS},
  journal={The Journal of Physical Chemistry C},
  volume={116},
  number={6},
  pages={3951--3955},
  year={2012},
  publisher={ACS Publications},
  doi={10.1021/jp210536m}}

@article{polley2023,
  title={Bottom-up growth of monolayer honeycomb SiC},
  author={Polley, CM and Fedderwitz, H and Balasubramanian, T and Zakharov, AA and Yakimova, Rositsa and B{\"a}cke, O and Ekman, J and Dash, SP and Kubatkin, S and Lara-Avila, S},
  journal={Physical Review Letters},
  volume={130},
  number={7},
  pages={076203},
  year={2023},
  publisher={APS},
  doi={10.1103/PhysRevLett.130.076203}}

@article{sahin2009,
  title={Monolayer honeycomb structures of group-IV elements and III-V binary compounds: First-principles calculations},
  author={{\c{S}}ahin, Hasan and Cahangirov, Seymur and Topsakal, Mehmet and Bekaroglu, E and Akturk, Ethem and Senger, R Tugrul and Ciraci, Salim},
  journal={Physical Review B—Condensed Matter and Materials Physics},
  volume={80},
  number={15},
  pages={155453},
  year={2009},
  publisher={APS},
  doi={10.1103/PhysRevB.80.155453}}

@article{ha2023,
  title={First-principles study of SiC and GeC monolayers with adsorbed non-metal atoms},
  author={Ha, Chu Viet and Ha, LT and Nguyen, Duy Khanh and Anh, Dang Tuan and Guerrero-Sanchez, J and Hoat, DM and others},
  journal={RSC advances},
  volume={13},
  number={22},
  pages={14879--14886},
  year={2023},
  publisher={Royal Society of Chemistry},
  doi={10.1039/D3RA01372C}}

@misc{supp,
    title={Link to Supplemental Material}}

@article{guo2024,
  title={Sliding ferroelectricity in kagome-B2X3 (X= S, Se, Te) bilayers},
  author={Guo, Yan-Dong and Meng, Rui-Jie and Hu, Xue-Qin and Lin, Li-Yan and Jiang, Yue and Yang, Ming-Yu and You, Yun and Zhang, Lan-Qi and Xu, Yi-Long and Yan, Xiao-Hong},
  journal={Applied Physics Letters},
  volume={124},
  number={15},
  year={2024},
  pages={152901},
  publisher={AIP Publishing},
  doi={10.1063/5.0198134}}

@article{bennett2023,
  title={Polar meron-antimeron networks in strained and twisted bilayers},
  author={Bennett, Daniel and Chaudhary, Gaurav and Slager, Robert-Jan and Bousquet, Eric and Ghosez, Philippe},
  journal={Nature Communications},
  volume={14},
  number={1},
  pages={1629},
  year={2023},
  publisher={Nature Publishing Group UK London},
  doi={10.1038/s41467-023-37337-8}}

@article{ooi2005,
  title={Electronic structure and bonding in hexagonal boron nitride},
  author={Ooi, N and Rairkar, A and Lindsley, L and Adams, JB},
  journal={Journal of Physics: Condensed Matter},
  volume={18},
  number={1},
  pages={97},
  year={2005},
  publisher={IOP Publishing},
  doi={10.1088/0953-8984/18/1/007}}

@article{smith2015,
  title={Interplay of octahedral rotations and lone pair ferroelectricity in CsPbF3},
  author={Smith, Eva H and Benedek, Nicole A and Fennie, Craig J},
  journal={Inorganic Chemistry},
  volume={54},
  number={17},
  pages={8536--8543},
  year={2015},
  publisher={ACS Publications},
  doi={10.1021/acs.inorgchem.5b01213}}

@article{shen2019,
  title={Role of Lone-Pairs in Driving Ferroelectricity of Perovskite Oxides: An Orbital Selective External Potential Study},
  author={Shen, Yang and Cai, Jia and Ding, Hang-Chen and Shen, Xin-Wei and Fang, Yue-Wen and Tong, Wen-Yi and Wan, Xian-Gang and Zhao, Qingbiao and Duan, Chun-Gang},
  journal={Advanced Theory and Simulations},
  volume={2},
  number={6},
  pages={1900029},
  year={2019},
  publisher={Wiley Online Library},
  doi={10.1002/adts.201900029}}

@article{jeong2021,
  title={Ferroelectric switching in GeTe through rotation of lone-pair electrons by Electric field-driven phase transition},
  author={Jeong, Kwangsik and Lee, Hyangsook and Lee, Changwoo and Wook, Lim Hyeon and Kim, Hyoungsub and Lee, Eunha and Cho, Mann-Ho},
  journal={Applied Materials Today},
  volume={24},
  pages={101122},
  year={2021},
  publisher={Elsevier},
  doi={10.1016/j.apmt.2021.101122}}

@article{gonze2020,
  title={The ABINIT project: Impact, environment and recent developments},
  author={Gonze, Xavier and Amadon, Bernard and Antonius, Gabriel and Arnardi, Fr{\'e}d{\'e}ric and Baguet, Lucas and Beuken, Jean-Michel and Bieder, Jordan and Bottin, Fran{\c{c}}ois and Bouchet, Johann and Bousquet, Eric and others},
  journal={Computer Physics Communications},
  volume={248},
  pages={107042},
  year={2020},
  publisher={Elsevier},
  doi={10.1016/j.cpc.2019.107042}}

@article{perdew1996,
  title={Generalized gradient approximation made simple},
  author={Perdew, John P and Burke, Kieron and Ernzerhof, Matthias},
  journal={Physical review letters},
  volume={77},
  number={18},
  pages={3865},
  year={1996},
  publisher={APS},
  doi={10.1103/PhysRevLett.77.3865}}

@article{vantroeye2016,
  title={Interatomic force constants including the DFT-D dispersion contribution},
  author={Van Troeye, Benoit and Torrent, Marc and Gonze, Xavier},
  journal={Physical Review B},
  volume={93},
  number={14},
  pages={144304},
  year={2016},
  publisher={APS},
  doi={10.1103/PhysRevB.93.144304}}

@article{mostofi2014,
  title={An updated version of wannier90: A tool for obtaining maximally-localised Wannier functions},
  author={Mostofi, Arash A and Yates, Jonathan R and Pizzi, Giovanni and Lee, Young-Su and Souza, Ivo and Vanderbilt, David and Marzari, Nicola},
  journal={Computer Physics Communications},
  volume={185},
  number={8},
  pages={2309--2310},
  year={2014},
  publisher={Elsevier},
  doi={10.1016/j.cpc.2014.05.003}}

@book{jackson,
  title={Classical Electrodynamics},
  author={Jackson, John David},
  year={1975},
  publisher={John Wiley \& Sons}}

@book{dresselhaus,
    title={Group Theory: Application to the Physics of Condensed Matter},
    author={Dresselhaus, M. S. and Dresselhaus, G. and Jorio, A.},
    year={2008},
    publisher={Springer}}

@article{tsipas2013,
  title={Evidence for graphite-like hexagonal AlN nanosheets epitaxially grown on single crystal Ag (111)},
  author={Tsipas, P and Kassavetis, S and Tsoutsou, D and Xenogiannopoulou, E and Golias, EGSA and Giamini, SA and Grazianetti, C and Chiappe, D and Molle, A and Fanciulli, M and others},
  journal={Applied Physics Letters},
  volume={103},
  number={25},
  pages={251605},
  year={2013},
  publisher={AIP Publishing},
  doi={10.1063/1.4851239}}

@article{chettri2021,
  title={Induced magnetic states upon electron--hole injection at B and N sites of hexagonal boron nitride bilayer: A density functional theory study},
  author={Chettri, B and Patra, PK and Lalmuanchhana and Lalhriatzuala and Verma, Swati and Rao, B Keshav and Verma, Mohan L and Thakur, Vishal and Kumar, Narender and Hieu, Nguyen N and others},
  journal={International Journal of Quantum Chemistry},
  volume={121},
  number={16},
  pages={e26680},
  year={2021},
  publisher={Wiley Online Library},
  doi={10.1002/qua.26680}}

@article{bacaksiz2015,
  title = {Hexagonal AlN: Dimensional-crossover-driven band-gap transition},
  author = {Bacaksiz, C. and Sahin, H. and Ozaydin, H. D. and Horzum, S. and Senger, R. T. and Peeters, F. M.},
  journal = {Phys. Rev. B},
  volume = {91},
  issue = {8},
  pages = {085430},
  numpages = {7},
  year = {2015},
  publisher = {American Physical Society},
  doi = {10.1103/PhysRevB.91.085430}}

@article{yang2024,
  title = {Light-Induced Complete Reversal of Ferroelectric Polarization in Sliding Ferroelectrics},
  author = {Yang, Qing and Meng, Sheng},
  journal = {Phys. Rev. Lett.},
  volume = {133},
  issue = {13},
  pages = {136902},
  numpages = {6},
  year = {2024},
  publisher = {American Physical Society},
  doi = {10.1103/PhysRevLett.133.136902}}

@article{capillas2007,
  title={Maximal symmetry transition paths for reconstructive phase transitions},
  author={Capillas, Cesar and Perez-Mato, JM and Aroyo, MI},
  journal={Journal of Physics: Condensed Matter},
  volume={19},
  number={27},
  pages={275203},
  year={2007},
  publisher={IOP Publishing},
  doi={10.1088/0953-8984/19/27/275203}}

@article{weng2025,
	title={Sliding-Induced Out-of-Plane Ferroelectricity of 2D MnPS3},
	author={Weng, Xiangchao and Gui, Jiabao and Chen, Wenjun and Tan, Junyang and Li, Shengnan and Tang, Lei and Zhang, Rongjie and Wei, Qiang and Xu, Jiachun and Teng, Changjiu and others},
	journal={Advanced Functional Materials},
	volume={35},
	number={42},
	pages={e03780},
	year={2025},
	publisher={Wiley Online Library},
	doi={10.1002/adfm.202503780}}

@article{liang2025,
	title={Multidirectional sliding ferroelectricity of rhombohedral-stacked InSe for reconfigurable photovoltaics and imaging applications},
	author={Liang, Qingrong and Zheng, Guozhong and Fan, Shuaiwei and Yang, Liu and Zheng, Shoujun},
	journal={Advanced Materials},
	volume={37},
	number={7},
	pages={2416117},
	year={2025},
	publisher={Wiley Online Library},
	doi={10.1002/adma.202416117}}

@article{jain2025,
	title={Theoretical Prediction of Sliding Ferroelectricity in MoSiGeN4 Bilayers},
	author={Jain, Ayushi and Bera, Chandan},
	journal={The Journal of Physical Chemistry C},
	volume={129},
	number={19},
	pages={8850--8856},
	year={2025},
	publisher={ACS Publications},
	doi={10.1021/acs.jpcc.5c01358}}

@article{marzari2012,
	author={Marzari, Nicola and Mostofi, Arash A and Yates, Jonathan R and Souza, Ivo and Vanderbilt, David},
	journal={Reviews of Modern Physics},
	volume={84},
	number={4},
	pages={1419--1475},
	year={2012},
	publisher={APS},
	doi={10.1103/RevModPhys.84.1419}}

@article{marzari2003,
	author={Marzari, Nicola and Souza, Ivo and Vanderbilt, David},
	journal={Psi-K Newsletter},
	volume={57},
	pages={129},
	year={2003}}

@book{altmann,
	title={Point-Group Theory Tables},
	author={Altmann, S. L. and Herzig, P.},
	year={1994},
	publisher={Oxford University Press}}

@book{toledano,
	title={Landau Theory Of Phase Transitions},
	author={Toledano, Pierre and Toledano, Jean-claude},
	year={1987},
	publisher={World Scientific Publishing Company}}

@book{landau,
	title={Statistical Physics},
	author={Landau, L. D. and Lifshitz, E. M.},
	year={1980},
	publisher={Elsevier}}

@article{zhi2025,
	title={Tracking Polarization Structure in PtTe2 at Three-Dimensional Atomic Resolution},
	author={Zhi, Aomiao and Chen, Zitao and Wang, Jianlin and Xu, Hongjun and Miao, Guangyao and Chen, Pan and Han, Hongbo and Cai, Chen and Li, Xiaomin and Yu, Guoqiang and others},
	journal={Chinese Physics Letters},
	volume={42},
	number={7},
	pages={070717},
	year={2025},
	publisher={Chinese Physical Society and IOP Publishing Ltd},
	doi={10.1088/0256-307X/42/7/070717}}

@article{kingsmith_1993, 
	title={Theory of Polarization of Crystalline Solids}, volume={47}, number={3}, journal={Physical Review B}, author={King-Smith, R.D. and Vanderbilt, David}, year={1993}, pages={1651-1654},
	doi={10.1103/PhysRevB.47.1651}}

@article{sai2009,
	title={Absence of critical thickness in an ultrathin improper ferroelectric film},
	author={Sai, Na and Fennie, Craig J and Demkov, Alexander A},
	journal={Physical review letters},
	volume={102},
	number={10},
	pages={107601},
	year={2009},
	publisher={APS},
	doi={10.1103/PhysRevLett.102.107601}}
\end{document}